\documentclass[11pt, oneside]{article}   	
\usepackage{geometry}                		
\usepackage{graphicx}				
\usepackage{amssymb}

\usepackage{graphicx}
\usepackage{placeins}
\usepackage{newtxmath}
\usepackage{natbib}
\usepackage{cleveref}
\usepackage{ulem}

\usepackage{newtxtext}
\usepackage{newtxmath}
\usepackage{natbib}
\usepackage{hyperref}
\usepackage{epstopdf, epsfig}
\usepackage[utf8]{inputenc}
\usepackage{amsmath}
\usepackage{color}
\usepackage{listings}
\usepackage{float}
\usepackage{xcolor}
\hypersetup{
    colorlinks = false,
    urlcolor   = blue,
    citecolor  = black,
}

\newcommand{\RomanNumeralCaps}[1]
\linenumbers
\newcommand{\aj}{\textcolor{black}} 

\title{\textbf{Indirect noise from weakly reacting inhomogeneities}}

\author{Animesh Jain$^1$, Andrea Giusti$^2$ \& Luca Magri$^{2,1,4,5}$\footnote{l.magri@imperial.ac.uk} \\ 
 \small{$^1$ Department of Engineering, University of Cambridge,
Cambridge CB2 1PZ, UK} \\
\small{$^2$ Aeronautics Department, Imperial College London, London, SW7 1AL, UK}\\
\small{$^3$ Department of Mechanical Engineering, Imperial College London, London, SW7 1AL, UK}\\
\small{$^4$ Department of Aerospace Engineering, Technion, Haifa, Israel (visiting)}\\
\small{$^5$ Isaac Newton Institute for Mathematical Sciences, Cambridge, CB3 0EH, UK (visiting)}
}

\date{}

\begin{document}
\maketitle

\begin{abstract}

{
Indirect noise is a significant contributor to aircraft engine noise, which needs to be minimized in the design of aircraft engines. 
Indirect noise is caused by the acceleration of flow inhomogeneities through a nozzle. High-fidelity simulations showed that  some flow inhomogeneities can be chemically reacting when they leave the combustor and enter the nozzle~\citep{giusti2019flow}. The state-of-art models, however, are limited to chemically non-reacting (frozen) flows. 
In this work, first, we propose a low-order model to predict  indirect noise in nozzle flows with  reacting  inhomogeneities. 
Second, we identify the physical sources of sound, which generate indirect noise via two physical mechanisms: (i)  chemical reaction generates compositional perturbations, thereby adding to compositional noise; and 
(ii)  exothermic  reaction generates entropy perturbations. 
Third, we numerically compute the nozzle transfer functions for different frequency ranges (Helmholtz numbers) and reaction rates (Damk\"{o}hler numbers) in subsonic flows with hydrogen and methane inhomogeneities. 
Fourth, we extend the model to supersonic flows. We find that  hydrogen inhomogeneities have a larger impact to indirect noise than methane inhomogeneities.  
Both the Damk\"{o}hler number and the Helmholtz number markedly influence the phase and magnitude of the transmitted and reflected waves, which affect sound generation and thermoacoustic stability. 
This work provides a physics-based low-order model, which can open new opportunities for predicting noise emissions and instabilities in aeronautical gas turbines with multi-physics flows.
}
\end{abstract}

\section{Introduction}

Aircraft engine manufacturers are striving to make aeroengines both cleaner and quieter to meet the stringent targets set by aviation advisory councils. On the one hand, to make gas-turbine combustors cleaner, flames typically operate in a lean regime. On the other hand, lean flames are sensitive to the turbulent environment of the combustor, which can result in significantly unsteady chemical and fluid dynamics, which, in turn, can add to sound generation via direct and indirect mechanisms. 
In the combustor, two categories of noise can be identified.  
Direct noise is caused by the unsteady flickering of the flame, which leads to a volumetric expansion and contraction of the flow, which, in turn, generates acoustic waves. If these acoustic waves propagate downstream of the combustor, these waves are perceived as noise~\citep[e.g.,][]{strahle1976noise}. Direct noise is a well-studied phenomenon both experimentally and numerically, as reviewed by~\citet{ihme2017combustion}. 
Indirect noise is caused by a different mechanism, that is, the acceleration of flow inhomogeneities through the nozzle downstream of the combustor~\citep[e.g.,][]{marble1977acoustic, strahle1976noise, cumpsty1979jet, williams1975generation, polifke2001constructive, morgans2016entropy, magri2016compositional}.
Indirect noise generated by  inhomogeneities in temperature is typically referred to as entropy noise \citep{cuadra1967acoustic, marble1977acoustic, bake2009entropy, duran2013solution}, whereas indirect noise generated by  inhomogeneities in the composition is referred to as compositional noise~\citep{magri2016compositional, magri2017indirect}.
Indirect noise can also affect the combustor's stability. If the acoustic waves generated from inhomogeneities reflect off the components downstream of the combustor, they can synchronize constructively with the heat released by the flames, which can lead to thermoacoustic instabilities~\citep{polifke2001constructive, goh2013influence, motheau2014mixed, morgans2016entropy}.
These instabilities can cause a reduction in the engine lifetime and may also lead to structural  failures \citep{dowling2015combustion}.
To understand and capture the key physical mechanisms of indirect noise in  low-order models for the preliminary design of aircraft engines, substantial research has been carried out. 
 \citet{marble1977acoustic} introduced a model to predict entropy noise for a compact nozzle flow, which was extended to non-compact nozzle flows by \citet{leyko2009comparison, duran2013solution}, among others. These models assumed the flow to have a homogeneous composition, which can be approximated as a single component flow. However, factors such as air-cooling and improper mixing can generate inhomogeneities in the flow composition.  
The models of homogeneous flows were generalized to multicomponent flows to calculate the indirect noise caused by compositional inhomogeneities in both compact and non-compact nozzle flows by \citet{magri2016compositional, magri2017indirect}. 
These studies were further generalized to flows with entropy generation due to flow dissipation \citep[e.g.,][]{de2019generalised, jain_magri_2022, jain2022compositional, jain2022sound, guzman2022scattering}.

In the the aforementioned studies the flow is assumed to be chemically frozen. 
\citet{giusti2019flow} performed a Large-Eddy-Simulation study on a realistic aero-engine combustor. They showed that some flow inhomogeneities are chemically reacting when they leave the combustor and enter the nozzle guide vane. The chemical reaction produces both changes in composition and entropy fluctuations.
\citet{patki2022entropy} modelled the entropy generation mechanisms from exothermic chemical reactions in laminar premixed flames, but physical mechanisms for sound generation were not investigated. 

\begin{figure}
\centering    
\includegraphics[width=1\textwidth]{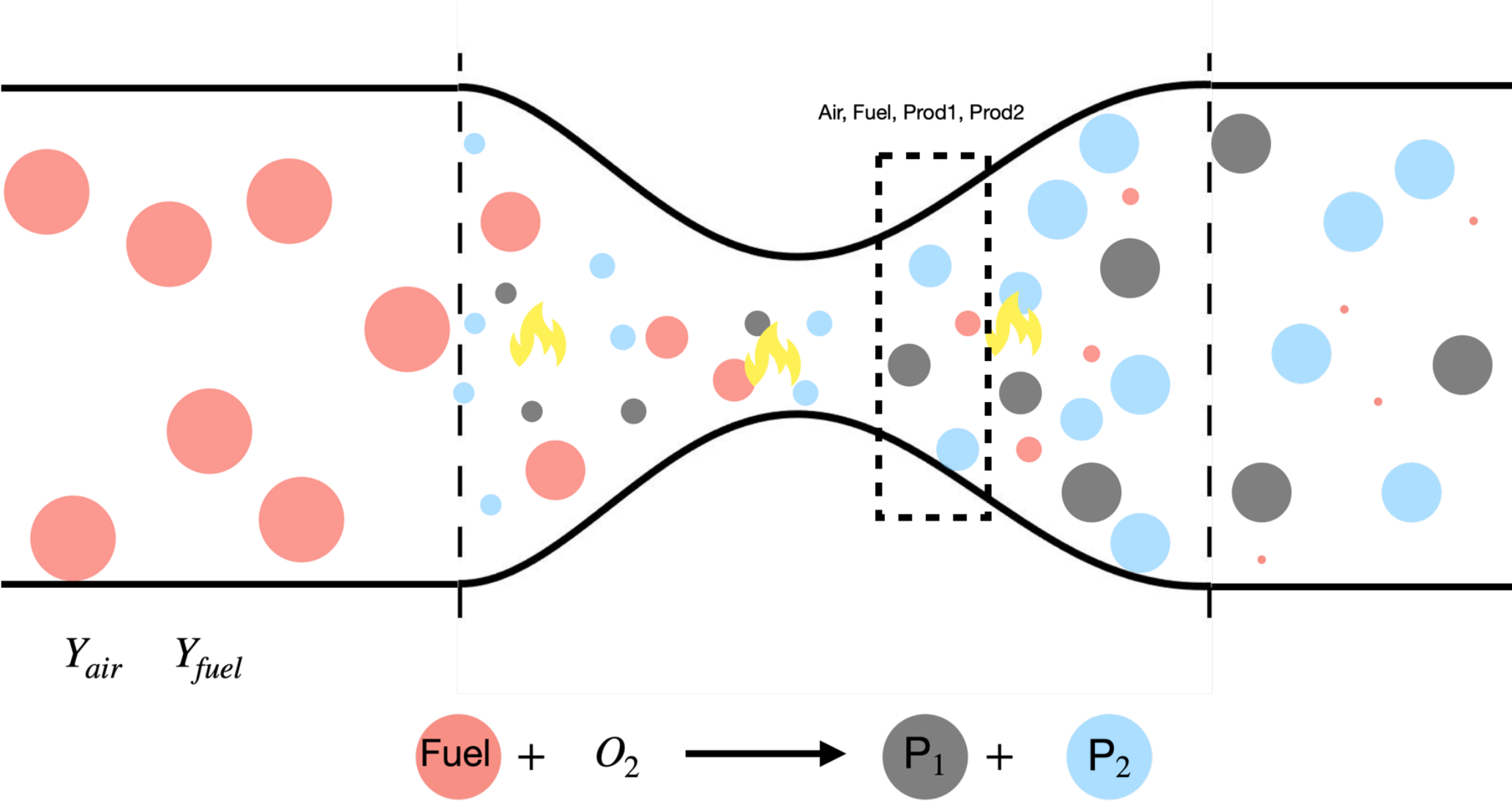}
\caption{Reacting flow schematic with an example of combustion of fuel with products $P_1$ and $P_2$.  
}
\label{fig:nozzle_react_schem}
\end{figure}

The overarching objective of this work is to derive from first principles the governing equations to model sound generation (indirect noise) in nozzles generated by {reacting flow inhomogeneities}. 
We will focus on weakly reacting flows, in which the volume of the reacting pocket of flow is small with respect to the volume of the mean flow (Figure \ref{fig:nozzle_react_schem}). 
We compute the effect of reacting inhomogeneities with hydrogen and methane fuels. 
Hydrogen is the potential future of energy and aviation because of its carbon-free reaction \citep{surer2018state, yusaf2022sustainable}. Because the combustion of hydrogen-based fuel blends  is achieved in a lean regime, they are susceptible to incomplete combustion as compared to hydrocarbon fuels \citep{hosseini2020overview}.  
Methane is the main component of natural gases, which have been widely used in stationary gas turbines for power generation~\citep{lefebvre2010gas}. 
Specifically, the goals of this work are to:
(i) propose a physics-based model to predict acoustics generated by the acceleration of chemically reacting inhomogeneities;
(ii) analyse the source of sound that generate indirect noise;
(iii) compute and analyse the transfer functions in subsonic  flows;
and (iv) generalize the model to supersonic nozzle flows. 
%
To achieve these goals, two convergent-divergent nozzles are numerically investigated. The paper is structured as follows.
Section~\ref{sec:mathmodel} introduces the physical model and identifies the sources of sound.
{Section~\ref{sec:chem_results} describes the chemistry model and the sources of sound.
Sections~\ref{sec:subsonicrect} and \ref{sec:supersonicreact} show the acoustic transfer functions in a subsonic and supersonic flow regime, respectively.
}
Conclusions end the paper.

\section{Mathematical Model}\label{sec:mathmodel}
In this section, we model a multi-component chemically reacting flow through a nozzle. 
We assume a multi-component flow that is dominated by quasi-one dimensional dynamics and has no viscous dissipation. 
The flow is adiabatic with negligible diffusion effects. 
We assume the chemical reactions to be nearly completed at the exit of the combustor with pockets of unburnt fuel. 
The reacting pockets occupy a small volume as compared to the mean flow.
With these assumptions, the conservation equations of mass, momentum, energy, and species are \citep{chiu1974theory}
\begin{align}
\frac{D\rho}{Dt} + \rho\frac{\partial u}{\partial x} + \frac{\rho u}{A}\frac{dA}{dx} &= {\Dot{S}_m}, \label{eq:dmdt_1}\\
    \frac{Du}{Dt} + \frac{1}{\rho}\frac{\partial p}{\partial x} &=  {\Dot{S}_M}, \label{eq:dmdt_132412}\\
    T\frac{Ds}{Dt} &= {\Dot{S}_s},\label{eq:dsdt_1}\\
    \frac{DY_i}{Dt} &= \Dot{S_Y}, \label{eq:2.4}
\end{align}
where
$\rho$ is the density,
$x$ is the axial distance,
$t$ is the time,
$u$ is the flow velocity,
$A$ is the area of the nozzle cross-section,
$p$ is the pressure,
$T$ is the temperature, and 
$s$ is the entropy ($s = \sum_{i=1}^N s_{i} Y_i$). The flow is assumed to consist of $N$ species with mass fractions $Y_i$.
The right-hand-side terms of \eqref{eq:dmdt_1}-\eqref{eq:2.4} are the source terms. We assume no mass generation, i.e., $\Dot{S}_m = 0$. The effects of body forces and friction are neglected, i.e., $\Dot{S}_M = 0$.
The chemical reaction in the flow adds to the energy generation through the entropy source term 
\begin{align}
    \Dot{S}_s &= - \sum_{i=1}^N \left(\frac{\mu_i}{W_i}\right) \frac{DY_i}{Dt},\label{eq:EST1}
\end{align}
where $\mu_i = W_i({\partial h}/{\partial Y_i}) = W_i({\partial g}/{\partial Y_i})$ is the chemical potential, $W_i$ is the molar mass, $h = \sum_{i=1}^N h_{i} Y_i$ is the specific enthalpy, and $g = h - T s$ is the specific Gibbs' energy.
The species source term is
\begin{align}
    \Dot{S_Y} &= \frac{1}{\rho}\dot\omega_i,
\end{align}
where $\dot\omega_i$ is the rate of production (or consumption) of species $i$ by chemical reaction.
These equations are closed by the Gibbs equation
\begin{align}   
T{ds} & = dh - \frac{dp}{\rho} - \sum_{i=1}^N \left(\frac{\mu_i}{W_i}\right) dY_i . \label{eq:2.5}
\end{align}
The gases are assumed to be ideal and calorically perfect, therefore 
\begin{align}\label{eq:handcp}
    &h = c_p(T - T^o), \ &c_p=\frac{\gamma}{\gamma - 1}R,
\end{align}
where $R = \mathcal{R}\sum_{i=1}^N Y_i/W_i$ is the specific gas constant of the mixture, $\mathcal{R}$ is the universal gas constant, $T^o$ is the temperature of the reference state, $\gamma = c_p/c_v$ is the heat capacity ratio, $c_p$ and $c_v$ are the specific heat capacities at constant pressure and constant volume, respectively, given by 
\begin{align}\label{eq:cpcv}
    &c_p = \sum_{i=1}^N c_{p,i} Y_i, \ 
    &c_v = \sum_{i=1}^N c_{v,i} Y_i.
\end{align}
Using \eqref{eq:2.5}, \eqref{eq:handcp}, $\mu_i/W_i = g_i = h_i - T s_i$ and the equation of state, $p=\rho R T$,  Gibbs equation becomes
\begin{align}\label{eq:ge_magri2017}
    \frac{ds}{c_p} = \frac{dp}{\gamma p} - \sum_{i=1}^N\left(\aleph_{1,i} + \psi_{1,i}\right)dY_i,
\end{align}
where
\begin{align}
    &\aleph_{1,i} = \frac{1}{\gamma - 1}\frac{d\log\gamma}{dY_i} + \frac{T^o}{T} \frac{d\log c_p}{dY_i},\label{eq:aleph1i}\\
    &\psi_{1,i} = \frac{1}{c_p T} \left(\frac{\mu_i}{W_i} - \Delta h^o_{f,i}\right)\label{eq:psi1i},
\end{align}
where 
$\Delta h^o_{f,i}$ is the enthalpy of formation of the $i$th species,
$\aleph_{1,i}$ and $\psi_{1,i}$ are the heat-capacity factor and chemical-potential functions, respectively.
Replacing \eqref{eq:cpcv} into  \eqref{eq:aleph1i}-\ref{eq:psi1i} yields 
\begin{align}\label{eq:dgammadyi1}
    \frac{d\log\gamma}{dY_{i}} &= \frac{1}{c_p}\sum_{j=1}^N c_{p_j}\frac{dY_j}{dY_i} - \frac{1}{c_v}\sum_{j=1}^N c_{v_j}\frac{dY_j}{dY_i}.
\end{align}
For $i\neq j$, in a chemically reacting flow, $dY_i/dY_j$  depends on the stoichiometric coefficients; however, in a chemically frozen flow, $dY_i/dY_j = 0$. {We exploit \eqref{eq:dgammadyi1} to physical interpret the results for a binary mixture in \S \ref{sec:chem_results}.} 
With chemical reactions, it is simpler to work in the mass fraction domain, as opposed to the mixture fraction domain  \citep[e.g.,][]{magri2017indirect, jain2022compositional}, which is the approach we take here. 
For completeness, the heat capacity factor and chemical potential functions defined in \citep{magri2017indirect} are related to the factor $\aleph_{1,i}$ and $\psi_{1,i}$ of this work by
\begin{align}
    &\aleph =  \sum_{i=1}^N\aleph_{1,i} \frac{dY_i}{dZ},\
    &\psi = \sum_{i=1}^N \psi_{1,i}\frac{dY_i}{dZ}.
\end{align}

\subsection{Linearization}\label{sec:linearization}
We model the acoustics as linear perturbations to a mean flow. 
For this, we decompose a generic flow variable as $(.)\rightarrow {\bar {(.)}}(x) + \epsilon(.)^{\prime}(x,t)$, where ${\bar {(.)}}(x)$ is the steady mean flow component, and $\epsilon(.)^{\prime}(x,t)$ is the first-order perturbation with $\epsilon \to 0$. On grouping the steady mean flow terms, the mean-flow equations are
\begin{align}
    &d(\bar\rho\bar u A) = 0,\
    &d(\bar p + \bar\rho\bar u^2) = 0.
\end{align}
{We assume that the mean flow has no dissipation, and consists of non-reacting constituents. The reaction rates of the mean flow quantities are zero, $\bar{\Dot{\omega_i}} = 0$. Therefore 
\begin{align}
    &d\bar s = 0,\
    &d\bar Y_i = 0. \label{eq:barYi}
\end{align}}
By grouping the first-order terms, we obtain the  equations that govern the acoustics and flow inhomogeneities 
\begin{align}
    &\frac{\bar D}{Dt}\frac{\rho^{\prime}}{\bar\rho} + \bar u\frac{\partial}{\partial x}\left(\frac{u^{\prime}}{\bar u}\right) = 0,\label{eq:mass1}\\
    &\frac{\bar D}{Dt}\left(\frac{u^{\prime}}{\bar u}\right) + \left(2\frac{u^{\prime}}{\bar u} + \frac{\rho^{\prime}}{\bar\rho} - \frac{p^{\prime}}{\bar p}\right) \left(\frac{\partial\bar u}{\partial x} \right) + \frac{1}{\bar\gamma}\left(\frac{\bar u}{\bar M^2}\right)\frac{\partial}{\partial x}\frac{p^{\prime}}{\bar p} = 0,\\
    \bar\rho&\frac{\bar D}{Dt}\left(\frac{s^\prime}{\bar c_p}\right) = -\frac{1}{\bar T\bar c_p}\sum_{i=1}^N \left(\frac{\bar\mu_i}{\bar W_i}\right)\dot\omega_i^\prime, \label{eq:linent_a} \\ 
    \bar\rho&\frac{\bar DY_i^\prime}{Dt} = {\Dot\omega_i^\prime}.\label{eq:species1}
\end{align}
where, $\bar D/ Dt = \partial/\partial t + \bar u \partial/\partial x$. To close the linear equations, we linearize the Gibbs equation \eqref{eq:ge_magri2017} and take the material derivative and combine the first-order terms to yield
\begin{align}\label{eq:ddtaurhodashrhobar}
    &\frac{\bar D}{Dt} \left(\frac{\rho^\prime}{\bar\rho}\right) =  \frac{\bar D}{Dt}\left(\frac{p^\prime}{\bar\gamma\bar p}\right) - \frac{\bar D}{Dt}\left(\frac{s^\prime}{\bar c_p} \right) - \sum_{i=1}^N\left(\bar\aleph_{1,i} + \bar\psi_{1,i}\right)\frac{\bar DY_i^\prime}{Dt} 
    - \frac{\gamma^\prime}{\bar\gamma}\frac{\bar D}{Dt}{\log\bar p^\frac{1}{\bar\gamma}},
\end{align} 
where $\gamma^\prime = \sum_{i=1}^N(d\gamma/dY_i) Y_i^\prime$ is the perturbation to the heat-capacity ratio, which  is  a function of the species mass fractions, $Y_i$, only.
The equation is integrated from an unperturbed state at {$t \to -\infty$} to yield the density fluctuation as
\begin{align}\label{eq:ge}
     \frac{\rho^\prime}{\bar\rho} &=  \frac{p^\prime}{\bar\gamma\bar p} - \frac{s^\prime}{\bar c_p} - \sum_{i=1}^N\left(\bar\aleph_{1,i} + \bar\psi_{1,i} + \bar\phi_{1,i}\right) Y_i^\prime - \bar\Theta,
\end{align} 
where
\begin{align}
    \bar\phi_{1,i} = \frac{d \log\gamma}{dY_i}\log\tilde p^{\frac{1}{\bar\gamma}},\label{eq:phi1i}
\end{align}
is the gamma-prime source of noise \citep{strahle1976noise, magri2017indirect}, and  $\Tilde{p}$ is the normalized pressure, {$\Tilde{p} = p/p_\text{ref}$}.
We identify the chemical-reaction noise factor
\begin{align}
    \bar\Theta = {\int_{-\infty}^{\tau}}\sum_{i=1}^N \phi_{1,i}\frac{DY_i^\prime}{Dt}dt,\label{eq:theta1i}
\end{align}
which physically depends on the variation of the heat-capacity ratio  and the reaction rate, $\Dot\omega^\prime$, through~\eqref{eq:species1}. If the flow is chemically frozen ($DY_i^\prime/Dt = 0$, thus, $\bar\Theta = 0$),  the density fluctuation~\eqref{eq:ge} tends to that of the non-reacting compositional noise model of \citet{magri2017indirect}.
On the one hand, if the flow is homogeneous, the density fluctuations depend only on pressure fluctuations. On the other hand, if the flow is inhomogeneous, the temperature and composition fluctuations also affect the density fluctuations. When these inhomogeneities accelerate through the nozzle, they contract and expand at a different rate than the mean flow, which generates momentum imbalance, hence, acoustic waves.
The factors that are responsible for the  density fluctuations generated by  compositional inhomogeneities ($\aleph_{i,1}, \psi_{i,1}, \phi_{i,1}, \Theta$)  are typically referred to as compositional noise factors \citep{magri2017indirect, jain2022compositional}.
The species with the largest mass fractions have the dominant effect on acoustics, as shown in \eqref{eq:ge}.

\subsubsection{Non-dimensional equations}\label{sec:goveq}

By combining \eqref{eq:ddtaurhodashrhobar}, \eqref{eq:ge} and \eqref{eq:mass1}-\eqref{eq:species1}, we eliminate the density fluctuation and obtain  
\begin{align}
    &\frac{\bar D}{D\tau}\left(\frac{p^\prime}{\bar\gamma\bar p}\right) + \Tilde{u} \frac{\partial}{\partial \eta}\left(\frac{u^{\prime}}{\bar u}\right) - \frac{\bar D}{D\tau}\left(\frac{s^\prime}{\bar c_p} \right)- \sum_{i=1}^N\left(\left(\bar\aleph_{1,i} + \bar\psi_{1,i}\right)\frac{\bar DY_i^\prime}{D\tau} + \tilde{u}\frac{d\log\bar p^\frac{1}{\bar\gamma}}{d\eta}\frac{d \log\gamma}{dY_i} Y_i^\prime\right) = 0, \label{eq:masslin}\\
    &\frac{\bar D}{D\tau}\left(\frac{u^{\prime}}{\bar u}\right) + \frac{1}{\bar\gamma}\left(\frac{\Tilde{u}}{\Tilde{M}^2}\right)\frac{\partial}{\partial\eta}\frac{p^{\prime}}{\bar p}+ \Bigg(2\frac{u^{\prime}}{\bar u} + \left(1 - \bar\gamma\right)\frac{p^\prime}{\bar\gamma\bar p} - \frac{s^\prime}{\bar c_p}  - \left(\bar\aleph_{1,i} + \bar\psi_{1,i} + \bar\phi_{1,i}\right) Y_i^\prime - \bar\Theta \Bigg)\left(\frac{\partial\Tilde{u}}{\partial\eta} \right) = 0,\label{eq:momlin}\\
    &\frac{\bar D}{D\tau}\left(\frac{s^\prime}{\bar c_p}\right) = -\frac{1}{\bar T\bar c_p}\sum_{i=1}^N \left(\frac{\bar\mu_i}{\bar W_i}\right)\tilde{\dot{\omega_i^\prime}},\label{eq:entlin}\\
    &\frac{\bar DY_i^{\prime}}{D\tau} = \tilde{\dot{\omega_i^\prime}}. \label{eq:specieslin}
\end{align}
The variables are normalized as
$t = \tau/f_a$,
$x = L\eta$ and
$\bar u = \tilde{u} c_\text{ref}$, where $f_a$ is the frequency of the impinging perturbations, $L$ is the length of the nozzle, $c_\text{ref}$ is the reference speed of the sound measured at the nozzle inlet. 
The material derivative is defined as 
$\bar D(.)/D\tau = \text{He} \partial(.)/\partial t + \tilde{u}\partial(.)/\partial \eta$, where $\text{He}$ is the Helmholtz number. The Helmholtz number is the ratio between the length of the nozzle and the wavelength of the impinging disturbances, $\text{He} = {f_a L}/{c_\text{ref}}$, which is a nondimensional number for the nozzle spatial extent. In a compact nozzle flow, $\text{He} = 0$.
The rate of production is non-dimensionalsied as $\tilde{\dot{\omega_i^\prime}} = ({L}/{c_\text{ref}})({{\dot{\omega_i^\prime}}}/{\bar\rho})$.
The momentum equation \eqref{eq:momlin} shows the mechanism of sound generation. The interaction between the flow acceleration, $\partial\tilde{u}/\partial\eta$, and the flow inhomogeneities appears as an acoustic source term. 
{ 
In particular, the interaction between the chemical-reaction noise factor, $\bar\Theta$, and flow acceleration, $\partial\tilde{u}/\partial\eta$, is a source that is not present in chemically frozen flows~\citep{magri2017indirect}.
}
The chemically reacting inhomogeneities have two main effects on the flow.
First, the composition of the inhomogeneity  changes, which results in different values of the terms of the compositional noise factor, which, in turn, depend on the mass fraction, $Y_i^\prime$ and properties of the chemical species produced or reacted. This can be observed in the Gibbs equation \eqref{eq:ge} and mass and momentum conservation equations \eqref{eq:masslin} and \eqref{eq:momlin}.
Second, {chemical reactions generate fluctuations in the entropy \eqref{eq:entlin}.} 
{By using the thermodynamic relationships $\mu_i/W_i = h_i - T s_i$ and $h_i = h_\text{sens} + h_\text{chem}$, the  right-hand-side term of \eqref{eq:entlin} can be cast as} 
\begin{align}
    &-\frac{1}{\bar T\bar c_p}\sum_{i=1}^N \left(\frac{\bar\mu_i}{\bar W_i}\right)\tilde{\dot{\omega_i^\prime}} = -\sum_{i=1}^N \bar\psi_{1,i} \tilde{\dot{\omega_i^\prime}} -  \frac{1}{\bar T\bar c_p} \sum_{i=1}^N\Delta h^o_{f,i}\tilde{\dot{\omega_i^\prime}},
\end{align}
where $\sum_{i=1}^N\Delta h^o_{f,i}\dot\omega_i^\prime = \mathcal{Q}\dot\omega_f^\prime$, and $\mathcal{Q}$ is the heat release per kilogram of reacted fuel. The entropy equation \eqref{eq:linent_a} can be written as
\begin{align}\label{eq:entDirectnoise}
    \frac{\bar D}{D\tau}\left(\frac{s^\prime}{\bar c_p}\right) = -\frac{\mathcal{Q}\tilde{\dot{\omega_f^\prime}}}{\bar T \bar c_p} -\sum_{i=1}^N \bar\psi_{1,i} \tilde{\dot{\omega_i^\prime}},
\end{align}
which shows that the entropy is generated because of (i) the heat released due to chemical reactions, $\mathcal{Q}$, and (ii) {the combined effect of the chemical reaction and chemical potential function ($\bar\psi_{1,i} \tilde{\dot{\omega_i^\prime}}$).} 
{If the flow is chemically frozen ($\tilde{\dot{\omega_i^\prime}}= 0$) the right-hand sides of \eqref{eq:entlin} and \eqref{eq:specieslin} are zero.}
In this limit, the set of equations \eqref{eq:masslin}-\eqref{eq:specieslin} tend to that of \citet{magri2017indirect}.

\subsubsection{Sources of noise}\label{sec:sourcesofnoise}
\begin{table}
  \begin{center}
\def~{\hphantom{0}}
  \begin{tabular}{c|c|c|c}
  \multicolumn{1}{c}{Source} &\multicolumn{1}{c}{Equation} &\multicolumn{1}{c}{Direct Noise} 
  &\multicolumn{1}{c}{Indirect Noise} \\
  \hline\hline
         $\sum_{i=1}^N \left(\bar\aleph_{1,i} + \bar\psi_{1,i} + \bar\phi_{1,i}\right) Y_i^\prime$ &\eqref{eq:momlin} & - &Compositional source\\[3pt]
         \hline
        $\bar\Theta$ &\eqref{eq:momlin}   & -  &Reacting compositional source\\
        \hline 
        $\hat{\mathcal{Q}} = \frac{\mathcal{Q}\tilde{\dot{\omega_f^\prime}}}{\bar T \bar c_p}$ &\eqref{eq:entDirectnoise}   & Heat release source  & - \\
        \hline
        $\hat{\psi}_{\tilde{\omega}} = \sum_{i=1}^N \bar\psi_{1,i} \tilde{\dot{\omega_i^\prime}}$ &\eqref{eq:entDirectnoise}   &Reacting chemical potential source  & - \\
        \hline
 \end{tabular}
  \caption{Sources of noise in a flow with weakly  reacting perturbations.}
  \label{tab:sources}
  \end{center}
\end{table}
The sources of noise can be identified from  \eqref{eq:masslin}-\eqref{eq:specieslin}.
{On the one hand,} the acoustic waves caused by the interaction of the {flow inhomogeneities}
with the flow acceleration contribute to indirect noise. On the other hand, the acoustic sources that do not depend on the flow acceleration are the direct noise sources.
The sources of noise in a flow with chemically reacting perturbations are summarised in Table \ref{tab:sources}. First, we identify two non-dimensional sources of direct noise by using the right-hand side of equation \eqref{eq:entDirectnoise}: heat release source, $\hat{\mathcal{Q}}$, and the reacting chemical potential source, $\hat{\psi}_{\tilde{\omega}}$ of direct noise. They act as (i) a monopole source of sound by appearing as a source term in the conservation of mass equation \eqref{eq:masslin}, and (ii) a source of entropy generation that affects the indirect noise through a change in fluctuation in the entropy, $s^\prime/\bar c_p$, in the momentum  equation \eqref{eq:momlin}.
The first direct-noise source is the heat release source
\begin{align}\label{eq:qhat}
    \hat{\mathcal{Q}} &= \frac{\mathcal{Q}\tilde{\dot{\omega_f^\prime}}}{\bar T \bar c_p},
\end{align}
in which $\mathcal{Q}$ depends on the stoichiometric ratios and the reaction chemistry. In an exothermic reaction, $\mathcal{Q}$ is positive, whereas, the rate or reaction, $\tilde{\dot{\omega_f^\prime}}$, is negative. Hence, \eqref{eq:entDirectnoise} shows that $\hat{\mathcal{Q}}$ results in an increase in the entropy fluctuations.  
The second direct-noise source is the reacting chemical potential source,
\begin{align}
    \hat{\psi}_{\tilde{\omega}} &= \sum_{i=1}^N \bar\psi_{1,i} \tilde{\dot{\omega_i^\prime}} {= \sum_{i=1}^N \bar\psi_{1,i} \frac{\bar DY_i^{\prime}}{D\tau}}, 
\end{align}
which is physically the interaction of the chemical potential and the rate of reaction of the species. {The chemical potential is the partial derivative of the Gibbs energy with respect to the number of moles at a constant pressure and temperature, $\mu_i = (\partial G/\partial n_i)_{p, T,n_{j\neq i}}$, which determines the direction in which species tend to migrate \citep{job2006chemical}.  Opposite signs of the chemical potential functions in a mixture correspond to opposite tendencies to mix \citep{jain2022compositional}.
In weakly reacting flows, the compositional inhomogeneity changes according to the rate of reaction through ${\bar DY_i^{\prime}}/{D\tau}$. Therefore, the reacting chemical potential source, $\hat{\psi}_{\tilde{\omega}}$, physically corresponds to the combined effect of the tendency to mix and the change in the species. 
Second, we identify two non-dimensional sources of indirect noise that add to  $s^\prime/\bar c_p$. The indirect-noise sources are the compositional noise source term, $\sum_{i=1}^N \left(\bar\aleph_{1,i} + \bar\psi_{1,i} + \bar\phi_{1,i}\right) Y_i^\prime$, which is similar to the compositonal noise source terms in a chemically frozen flow \citep{magri2017indirect}, and 
the reaction compositional noise source, $\bar\Theta$, as discussed in $\S~$\ref{sec:linearization}.
We compute the effect of these sources on the acoustic transfer functions in $\S\S~$\ref{sec:subsonicrect} and \ref{sec:supersonicreact}.

\subsubsection{Numerical solution}\label{sec:numsol}
The equations are solved numerically by Fourier transforming \eqref{eq:masslin} - \eqref{eq:specieslin} with the decomposition $\mathbf{q}(\tau,\eta) = \hat{\mathbf{{q}}}({\eta})\exp(2\pi i \tau)$. The primitive variables are expressed as travelling waves as $\pi^\pm = 0.5\left[{p^\prime}/{(\bar \gamma \bar p)} \pm {u^\prime}/{\bar u} \right]$;  the entropy inhomogeneity as the advected quantity
   $ \sigma = {s^\prime}/{\bar c_p}$; and the compositional inhomogeneity as the advected quantity $\xi = Y_f^\prime$.
The quantities of interest are the acoustic transfer functions, which are the reflection and transmission coefficients, respectively 
\begin{align}
    &R_\xi={\pi_1^-}/{\xi},\ 
    &T_\xi ={\pi_2^+}/{\xi}. \label{eq:frjfr3ij4}
\end{align}
The transfer functions are complex, therefore, they have a phase and a magnitude. In compact nozzles ($\mathrm{He} = 0$) the phase is zero.
The mass fraction of the products is assumed to be zero at the inlet. Chemically, the mass fraction of the fuel and product change along the nozzle depending on the rate of reaction, $\Dot\omega_i$.

\section{Chemistry models and  sources of sound}\label{sec:chem_results}

{In the general formulation presented in $\S~$\ref{sec:mathmodel}, the rate of production $
\dot\omega^\prime$ is not prescribed. Therefore, it can be obtained from detailed chemistry calculations.  
To reduce the complexity of the model, we prescribe a chemistry model for the rate of production, which closes the equations. }

\subsection{Reaction chemistry}\label{sec:rxnchem}
\FloatBarrier
{We assume that the chemiacl reaction is single-step and irreversible
\begin{align}
    {a}(\text{fuel}) + {b}\left(\text{air}\right) \rightarrow {c}(\text{products}),
\end{align}
where, $a, b$ and $c$ are the stoichiometric coefficients. 
{The rate of production of the fuel can be prescribed as}~
\citep{lieuwen2012unsteady}
\begin{align}\label{eq:omegafdot}
    \dot\omega^\prime_f = -\mathcal{A} \bar\rho Y_f^\prime,
\end{align}
where the subscript $f$ stands for fuel; and $\mathcal{A} = A_1 \exp{({-E_a}/{\mathcal{R}_u T})}$, where $E_a$ is the activation energy and $A_1$ is the pre-exponential coefficient. {For simplicity,} we assume $\mathcal{A}$ to be a constant in a flow. 
(Note that equation \eqref{eq:omegafdot} is only a function of $Y_f^\prime$ because, upon linearization, $Y_f^\prime \bar Y_\text{air} \gg Y_\text{air}^\prime \bar Y_f$. This is because a small amount of inhomogeneities of fuel is assumed to enter the nozzle, $\bar Y_\text{air} \gg \bar Y_\text{f}$, and the mean flow is not reacting~\eqref{eq:barYi}. Therefore,  $\bar Y_\text{air}$ is approximately constant and can be included in the coefficient, $\mathcal{A}$.)
{ 
We assume that a  fuel inhomogeneity is forced over the mean flow. As a result of the chemical reactions, we observe a decaying amplitude of the fluctuations of the fuel  (as shown in Figure \ref{fig:YfDaeta}).
Physically, the  fuel inhomogeneities become smaller and generate new gas pockets of products.}

\subsection{Damk\"{o}hler number}
The Damk\"{o}hler number is defined as 
\begin{align}\label{eq:Dadef}
    \text{Da} = \frac{\tau_\text{flow}}{\tau_\text{chem}} = \frac{L/c_\text{ref}}{1/\mathcal{A}},
\end{align}
where $\tau_\text{flow}$ is the characteristic hydrodynamic time scale, and $\tau_\text{chem}$ is the chemical reaction time scale. Therefore, the rate of production \eqref{eq:omegafdot} can be conveniently expressed as 
\begin{align}\label{eq:omegayf}
    \tilde{\dot{\omega_f^\prime}}  = \text{Da} Y_f^\prime.
\end{align}

From stoichiometry,
\begin{align}\label{eq:stoich}
    \frac{\dot\omega_f^\prime/ W_f}{- a} = \frac{\dot\omega_{\text{air}}^\prime/ W_{\text{air}}}{- b} = \frac{\dot\omega_{\text{prod}}^\prime/ W_{\text{prod}}}{c},
\end{align}
{which relates the rates of production of different species with the rate of production of the fuel.}
We can write the linearized entropy and species equation {as functions of the Damk\"{o}hler number and mass fraction of the fuel, $Y_f^\prime$, by using \eqref{eq:omegayf} and \eqref{eq:stoich}}, as 
\begin{align}
    &\frac{\bar D}{D\tau}\left(\frac{s^\prime}{\bar c_p}\right) = -\frac{a_{i,f}\text{Da}}{\bar T\bar c_p}\sum_{i=1}^N \left(\frac{\bar\mu_i}{\bar W_i}\right)Y_f^\prime,\label{eq:entlin_b}\\
    &\frac{\bar DY_i^{\prime}}{D\tau} = a_{i,f}\text{Da} ~Y_f^\prime, \label{eq:specieslin_b}
\end{align}
where $a_{i,f}$ is the ratio of products of stoichiometric coefficients and molecular weight of species i and fuel (e.g., $a_{\text{prod},f} = -c W_\text{prod}/(a W_{f})$).
If the reaction time scale is large ($\mathrm{Da} \ll 1$), the flow can be treated as chemically frozen; whereas, if the flow time scale is large ($\mathrm{Da} \gg 1$), the reaction occurs nearly instantaneously at the inlet of the nozzle. This can be observed in Figure \ref{fig:YfDaeta} (a). 
The magnitude remains almost constant for small Damk\"{o}hler numbers (Da $< 0.001$). However, for  large Damk\"{o}hler numbers, the magnitude of the fuel fluctuations rapidly decreases near the inlet of the nozzle and remains approximately zero thereafter (Da $> 10$ in Figure \ref{fig:YfDaeta} (a)). The flow can  be approximated as chemically frozen after the reaction is completed.
{In \S\S \ref{sec:subsonicrect} and \ref{sec:supersonicreact}, we choose a Damk\"{o}hler number that represents a general case in which the reaction continues throughout the nozzle flow (i.e., Da $\not\ll 1$ and Da $\not \gg 1$). Hence, we choose a Damk\"{o}hler number, Da = $0.05$, to show examples of general behaviour in a chemically reacting flow.}
Figure \ref{fig:YfDaeta} (b) {shows the fluctuations in the fuel and product mass fractions  in a reacting flow with Damk\"{o}hler number Da $= 0.05$, and Helmholtz number He $= 0.5$.}
For numerical analysis and computation of the acoustic transfer functions, we impose a unit amplitude inhomogeneity wave of the fuel at the nozzle inlet. In Figure \ref{fig:YfDaeta} (b), we observe that, due to the chemical reaction, the amplitude of the fluctuations decreases with the nozzle spatial location. At the same time,  product mass  is generated and the amplitude increases along the nozzle. In the chemically frozen flow, the amplitude of the fuel mass fraction  is constant and equal to $1$. (In the figures, a negative value of fluctuations $Y^\prime$ does not imply a negative mass fraction. This is because we linearize the mass fraction as $Y \rightarrow {\bar {Y}}(x) + \epsilon Y^{\prime}(x,t)$ in $\S~$\ref{sec:linearization}. 
)     
\begin{figure}
\centering    
\includegraphics[width=1\textwidth]{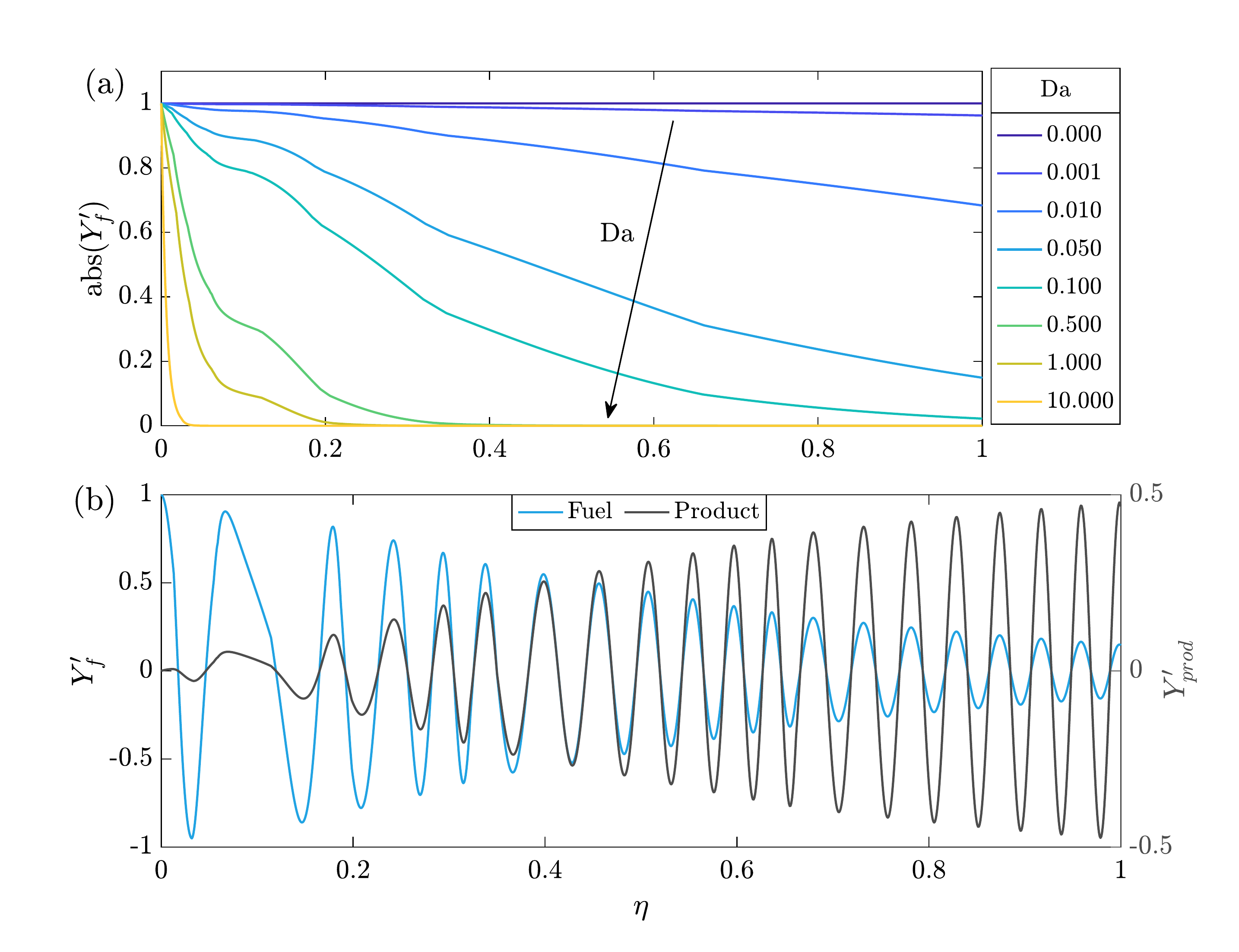}
\caption{(a) Absolute value of the fluctuations in the mass fraction of fuel for different Damk\"{o}hler numbers, Da, and Helmholtz numbers, He = 0.5.
(b) Fluctuations in the mass fraction of fuel (left) and products (right) for Da = 0.05 and He = 0.5, in a subsonic flow.
The horizontal axis shows the non-dimensionalized nozzle location, where $\eta = 0$ is the nozzle inlet and $\eta = 1$ is the outlet.}
\label{fig:YfDaeta}
\end{figure}
\FloatBarrier

{
\subsection{Sources of noise}
\FloatBarrier
Under the assumptions made in $\S~$\ref{sec:rxnchem}, we have three components in the flow: air, fuel, and products (Figure \ref{fig:nozzle_react_schem}), whose mass fractions fulfil the conservation of species
\begin{align}
    Y_{\text{air}} + Y_{\text{fuel}} + Y_{\text{products}} = 1.
\end{align}
On linearizing
\begin{align}
    Y_{\text{air}}^\prime = \underbrace{1 - \bar Y_{\text{air}} - \bar Y_{\text{fuel}}}_{\delta} - \left(Y_{\text{fuel}}^\prime + Y_{\text{products}}^\prime\right).
\end{align}
We assume {$\bar Y_{\text{air}} \gg \bar Y_{\text{fuel}}$ and $\bar Y_{\text{air}} \approx 1$ ($\S~$\ref{sec:rxnchem}), hence,} $\delta \to 0$. From \eqref{eq:aleph1i} and $\aleph_1 = \sum_{i=1}^N \aleph_{1,i} Y_i$, the heat capacity factor is 
\begin{align}\label{eq:alephreactass}
    \aleph_1 &= \frac{1}{\bar\gamma - 1} \left[\left(\frac{d\log\gamma}{dY_{\text{f}}} - \frac{d\log\gamma}{dY_{\text{air}}}\right) Y^\prime_{\text{f}} + \left(\frac{d\log\gamma}{dY_{\text{prod}}} - \frac{d\log\gamma}{dY_{\text{air}}}\right) Y^\prime_{\text{prod}}\right].
\end{align}
Similarly, the chemical potential function can be written as
\begin{align}\label{eq:psireactass}
    \psi_1 &= \left(\bar\psi_\text{f} - \bar\psi_\text{air}\right)Y_\text{f}^\prime + \left(\bar\psi_{\text{prod}} - \bar\psi_\text{air}\right)Y_{\text{prod}}^\prime.
\end{align}
The expressions \eqref{eq:alephreactass} and \eqref{eq:psireactass} {tend to} those of individual binary mixtures of fuel and products with air \citep{jain2022compositional}. 
We can conclude that, together with entropy generation (\eqref{eq:entDirectnoise}), another effect of chemical reactions on indirect noise is to change the composition of the inhomogeneities, which, in turn, changes the compositional noise factors. 
The compositional noise factors as functions of the properties of the fuel, products, and their mass fractions.
{The gamma-prime noise term, $\phi_1 = \sum_{i=1}^N \phi_{1,i} Y_i$, can be written as} 
\begin{align}\label{eq:phireactass}
    \phi_1 &= \log\bar p^{\frac{1}{\bar\gamma}} \left(\bar\gamma - 1\right) \aleph_1.
\end{align}
Compositional noise is not the same as that of a flow with a binary mixture of chemically frozen species. This is because, in a flow with chemical reactions, $dY_i/dY_j$ in  \eqref{eq:dgammadyi1} is a function of the molecular weights and stoichiometric coefficients of the chemical reaction.
}
The reacting compositional noise source of indirect noise, $\bar\Theta$, becomes 
\begin{align}
    \bar\Theta &= \int_{\tau=-\infty}^\tau \frac{\log\bar p}{\bar\gamma} \sum_{j=1}^N\frac{d\log\gamma}{dY_j}a_{j,f}\text{Da} ~Y_f^\prime d\tau, 
\end{align}
which, from $\S~$\ref{sec:numsol}, becomes 
\begin{align}
    \bar\Theta & = -\text{Da}\frac{\log\bar p}{\bar\gamma} \sum_{j=1}^N\frac{d\log\gamma}{dY_j}a_{j,f}\hat{Y_f^\prime}(\eta)\frac{e^{2\pi i \tau}}{2\pi} i .
\end{align}
{The reacting compositional noise source depends on the rate of production of the fuel, mean flow properties, and  time.}
In the limit of a chemically frozen flow, $\mathrm{Da} \ll 1$ (\eqref{eq:entlin_b} and \eqref{eq:specieslin_b}), and binary mixtures,  $dY_i/dY_j = -1$ in \eqref{eq:dgammadyi1}, the compositional noise factors (\eqref{eq:alephreactass}, \eqref{eq:psireactass}, and \eqref{eq:phireactass}) tend exactly to the 
 chemically frozen factors of \citet{magri2017indirect, jain2022compositional}.

\FloatBarrier

\subsection{Methane and Hydrogen compositional inhomogeneities}
We investigate the role of chemical reactions on indirect noise for inhomogeneities of methane and hydrogen, which are two fuels of interest to energy conversion and  aeronautical propulsion. 

\subsubsection{Methane reaction}
The chemical reaction of methane is 
\begin{align}\label{eq:ch4comb}
    \text{CH}_4 + 2\text{O}_2 \rightarrow \text{CO}_2 + 2\text{H}_2\text{O}.
\end{align} 
The heat released per kilogram of methane burned is $\mathcal{Q} = 50100$ kJ/kg \citep{poinsot2005theoretical}. 
The reacting compositional noise factors can be calculated from \eqref{eq:aleph1i}, \eqref{eq:psi1i}, \eqref{eq:phi1i} and \eqref{eq:theta1i}. The factor ${d\log\gamma}/{dY_{i}}$ for methane reaction can be calculated as 
\begin{align}
    \frac{d\log\gamma}{dY_{\text{CH}_4}} &= \frac{1}{c_p}\left(c_{p_{\text{CH}_4}} - c_{p_{{\text{CO}_2}}} \frac{(W)_{{\text{CO}_2}}}{(W)_{\text{CH}_4}} - c_{p_{\text{H}_2\text{O}}} \frac{2(W)_{\text{H}_2\text{O}}}{(W)_{\text{CH}_4}}\right)\ldots\nonumber\\
    &\ldots-\frac{1}{c_v}\left(c_{v_{\text{CH}_4}} - c_{v_{{\text{CO}_2}}} \frac{(W)_{{\text{CO}_2}}}{(W)_{\text{CH}_4}} - c_{v_{\text{H}_2\text{O}}} \frac{2(W)_{\text{H}_2\text{O}}}{(W)_{\text{CH}_4}}\right).\label{eq:dloggammadyi_ch4}
\end{align}
The detailed calculation is shown in {the supplementary material}.

\subsubsection{Hydrogen reaction}\label{sec:h2combust}
\FloatBarrier
The chemical reaction of hydrogen is 
\begin{align}\label{eq:h2comb}
    \text{H}_2 + \frac{1}{2}\text{O}_2 \rightarrow  \text{H}_2\text{O}.
\end{align}
For the analysis, we do not consider the formation of NO$_x$. The heat released per kilogram of hydrogen burned is $\mathcal{Q} = 120500$ kJ/kg \citep{poinsot2005theoretical}. 
The factor ${d\log\gamma}/{dY_{i}}$ for hydrogen reaction can be calculated as 
\begin{align}
    \frac{d\log\gamma}{dY_{\text{H}_2}} &= \frac{1}{c_p}\left(c_{p_{\text{H}_2}} - c_{p_{\text{H}_2\text{O}}} \frac{(W)_{\text{H}_2\text{O}}}{(W)_{\text{H}_2}}\right) -\frac{1}{c_v}\left(c_{v_{\text{H}_2}} - c_{v_{\text{H}_2\text{O}}} \frac{(W)_{\text{H}_2\text{O}}}{(W)_{\text{H}_2}}\right).\label{eq:dloggammadyi_h2}
\end{align} 
Equations \eqref{eq:dloggammadyi_ch4} and \eqref{eq:dloggammadyi_h2} show that ${d\log\gamma}/{dY_{i}}$ is constant for the assumptions made in this section.
Under the assumptions of single-step irreversible reaction, there are two main differences between the two reactions \eqref{eq:ch4comb} and \eqref{eq:h2comb}.
First, the heat of reaction (kJ/kg) in hydrogen is twice as large as that of methane.
Second, there is the formation of carbon dioxide in methane reaction. 
The magnitudes of the transfer function depend on the sources of noise, which are functions of the reaction chemistry, properties of fuel, and products. Depending on the Damk\"{o}hler number, the fuel converts to the products, and the sources of noise tend to exhibit properties of the products.
{The heat capacity factor, $\bar\aleph_1$, of the hydrogen-air mixture is approximately ten times larger than that of the methane-air mixture ($\eta = 0$ in Figure \ref{fig:indirectnoisefactors_subsonic} (a - b) (i)). However, the heat capacity factor, $\bar\aleph_1$ is negative for binary mixtures of both $\text{H}_2\text{O}$ and ${\text{CO}_2}$ with air with a comparable magnitude approximately half of that of methane and air mixture.
Similarly, the chemical potential indirect noise factor, $\bar\psi_1$, of the hydrogen-air mixture is approximately ten times larger in magnitude than the methane-air mixture. Both of them are negative. However,  $\bar\psi_1$ of ${\text{CO}_2}$-air mixture is positive with a magnitude approximately half of that of the methane-air mixture, and that of $\text{H}_2\text{O}$-air mixture is positive, but with a magnitude comparable to that of the methane-air mixture.
The gamma-prime noise factor, $\bar\phi_1$ is a function of the mean flow. The spatial variation depends on the variation of pressure across the flow.
}
$1$ kg of methane \eqref{eq:ch4comb}  produces $2.75$ kg of ${\text{CO}_2}$ and $2.25$ kg of $\text{H}_2\text{O}$. However, $1$ kg of hydrogen \eqref{eq:h2comb}  produces $9$ kg of $\text{H}_2\text{O}$, which is four times larger than that of methane. These properties affect the noise sources and, in turn, the transfer functions. In \S\S ~\ref{sec:subsonicrect}, \ref{sec:supersonicreact}, we analyse the effect of the Damk\"{o}hler number on the sources of noise, and the acoustic transfer functions in subsonic and supersonic regimes.

\section{Acoustic transfer functions {in subsonic flows}}\label{sec:subsonicrect}
\FloatBarrier
\begin{figure}
\centering    
\includegraphics[width=0.6\textwidth]{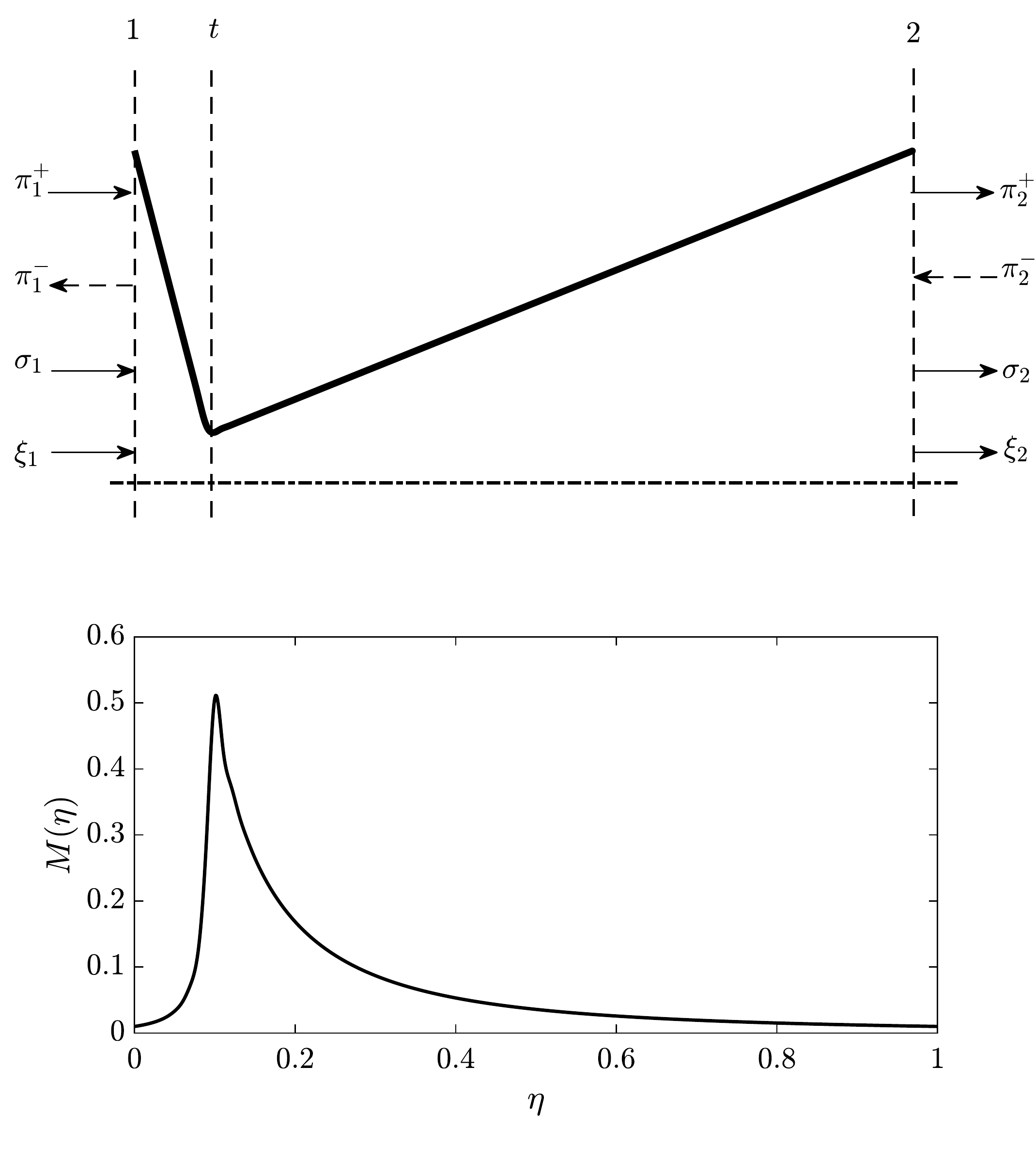}
\caption{(Top) Cambridge Wave Generator Nozzle profile. (Bottom) spatial variation of the Mach number.}
\label{fig:subsonicnozz_mach}
\end{figure}
In a subsonic regime, we investigate  two nozzle profiles. First, a converging-diverging nozzle similar to that of the experimental setup of the Cambridge Entropy Generator Rig \citep{de2017measurements} with a throat diameter of $6.6$ mm (Figure \ref{fig:subsonicnozz_mach}). This nozzle will be referred to as {CWG nozzle}. The inlet and outlet diameters are $46.2$ mm; the length of the converging and diverging sections are $24$ mm and $230$ mm, respectively; and the vena contracta factor is $\Gamma = 0.89$.
Second, in order to compare the results of the subsonic flow with the supersonic flow regime {($\S~$\ref{sec:supersonicreact})},
we use the nozzle with a linear-velocity profile in the subsonic regime \citep{magri2017indirect} (Figure \ref{fig:linvelnozz_mach}) {named `lin-vel nozzle', in this work}.  The nozzle profile and variation of the Mach number are shown in Figure \ref{fig:linvelnozz_mach}.
In both cases, the inlet pressure and temperature are $10^5$ Pa and $1000$ K, respectively. 
Additionally, we assume the flow to be composed of air with small pockets of fuel with $\bar\gamma = 1.4$. 
\\

\subsection{Sources of noise}\label{sec:son_subsonic}
\begin{figure}
\centering    
\includegraphics[width=1\textwidth]{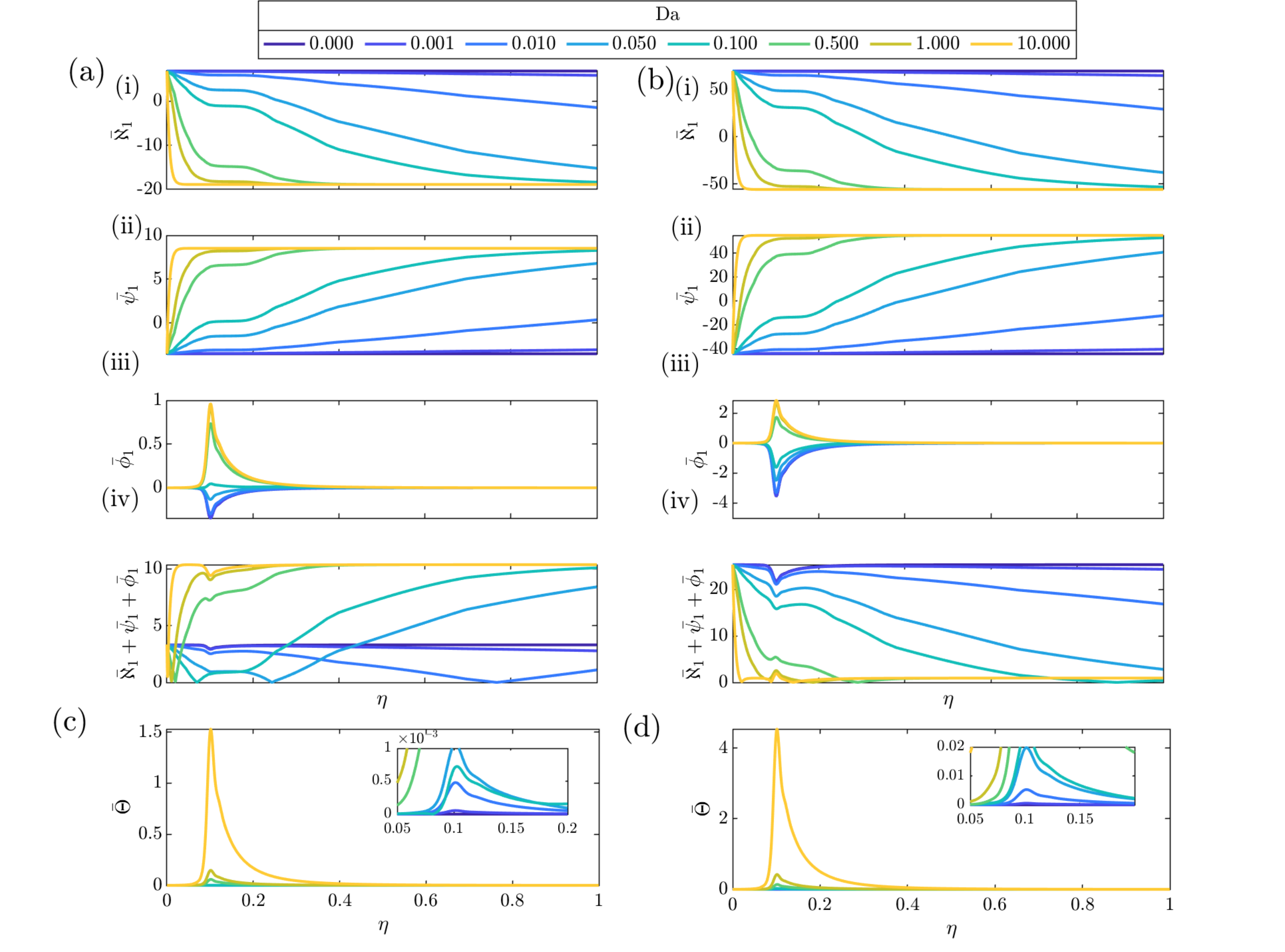}
\caption{Indirect noise factors. (a, b) Compositional indirect noise sources, (c, d) reacting compositional noise source  \aj{(Inset: close-up around nozzle throat)} in a subsonic flow ({CWG nozzle}) with He = 0.5 for (a, c) methane fuel and (b, d) hydrogen fuel.}
\label{fig:indirectnoisefactors_subsonic}
\end{figure}
\begin{figure}
\centering    
\includegraphics[width=1\textwidth]{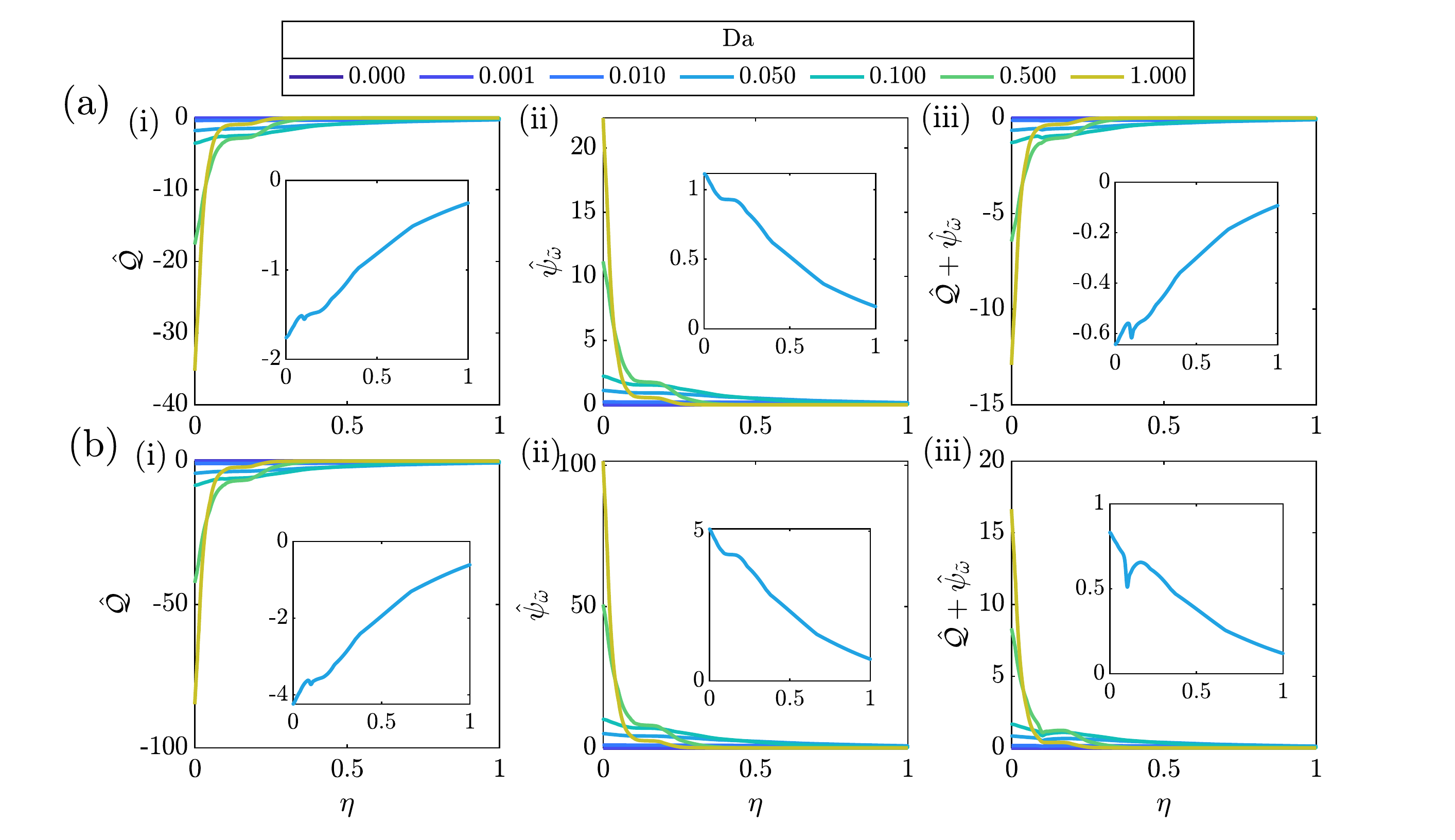}
\caption{Direct noise factors. (i) Heat noise source, (ii) reacting compositional noise source, (iii) total reacting direct noise source for (a) methane and (b) hydrogen fuel in a subsonic flow ({CWG nozzle}) with He = 0.5. \aj{Inset: Direct noise factors for $\mathrm{Da} = 0.05$.}}
\label{fig:directnoisefactors_subsonic}
\end{figure}
\begin{figure}
\centering    
\includegraphics[width=1\textwidth]{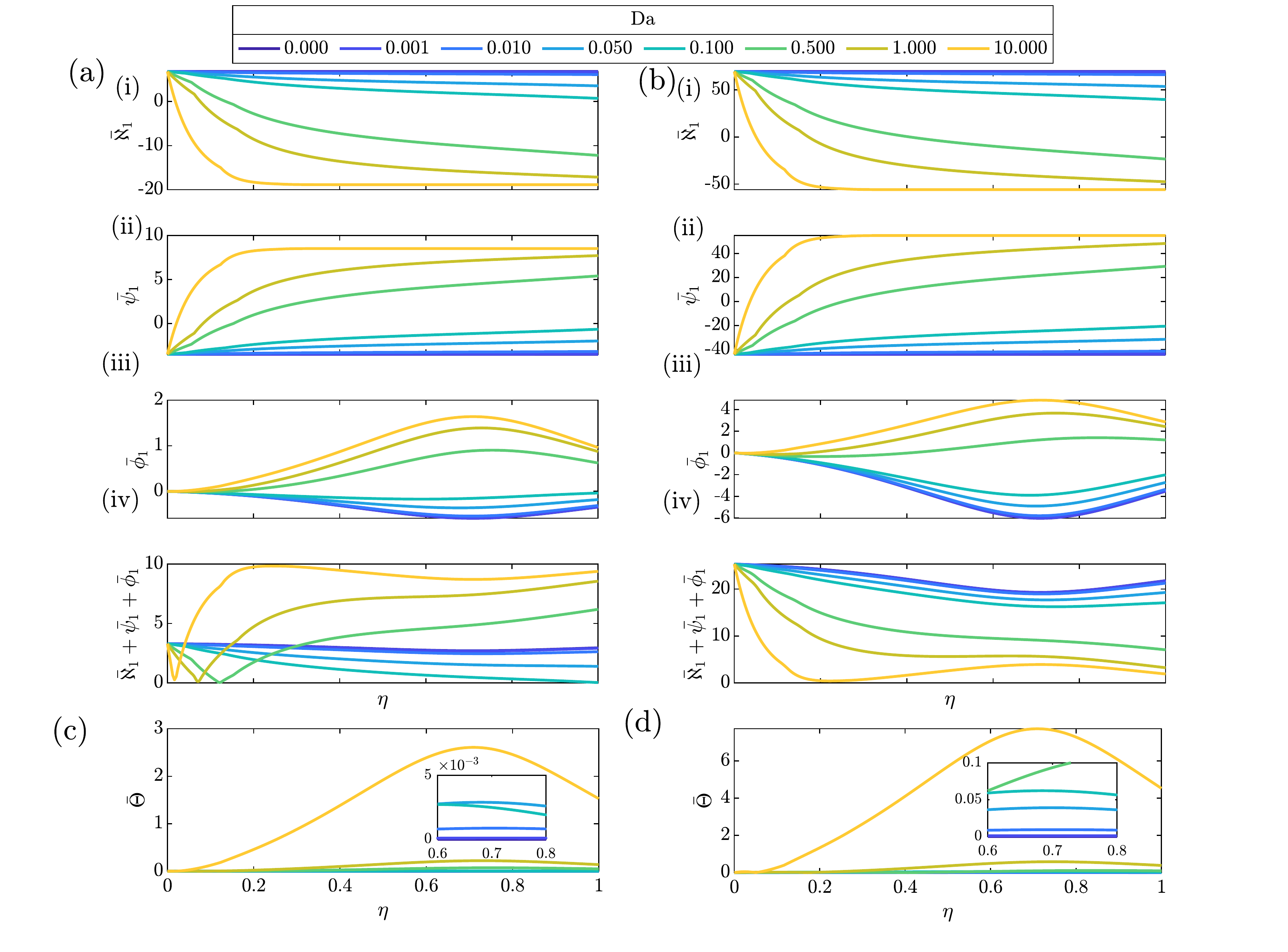}
\caption{
Same quantities as Figure \ref{fig:indirectnoisefactors_subsonic} for the lin-vel nozzle.
}
\label{fig:indirectnoisefactors_subsonic_linvelnozz}
\end{figure}
\begin{figure}
\centering    
\includegraphics[width=1\textwidth]{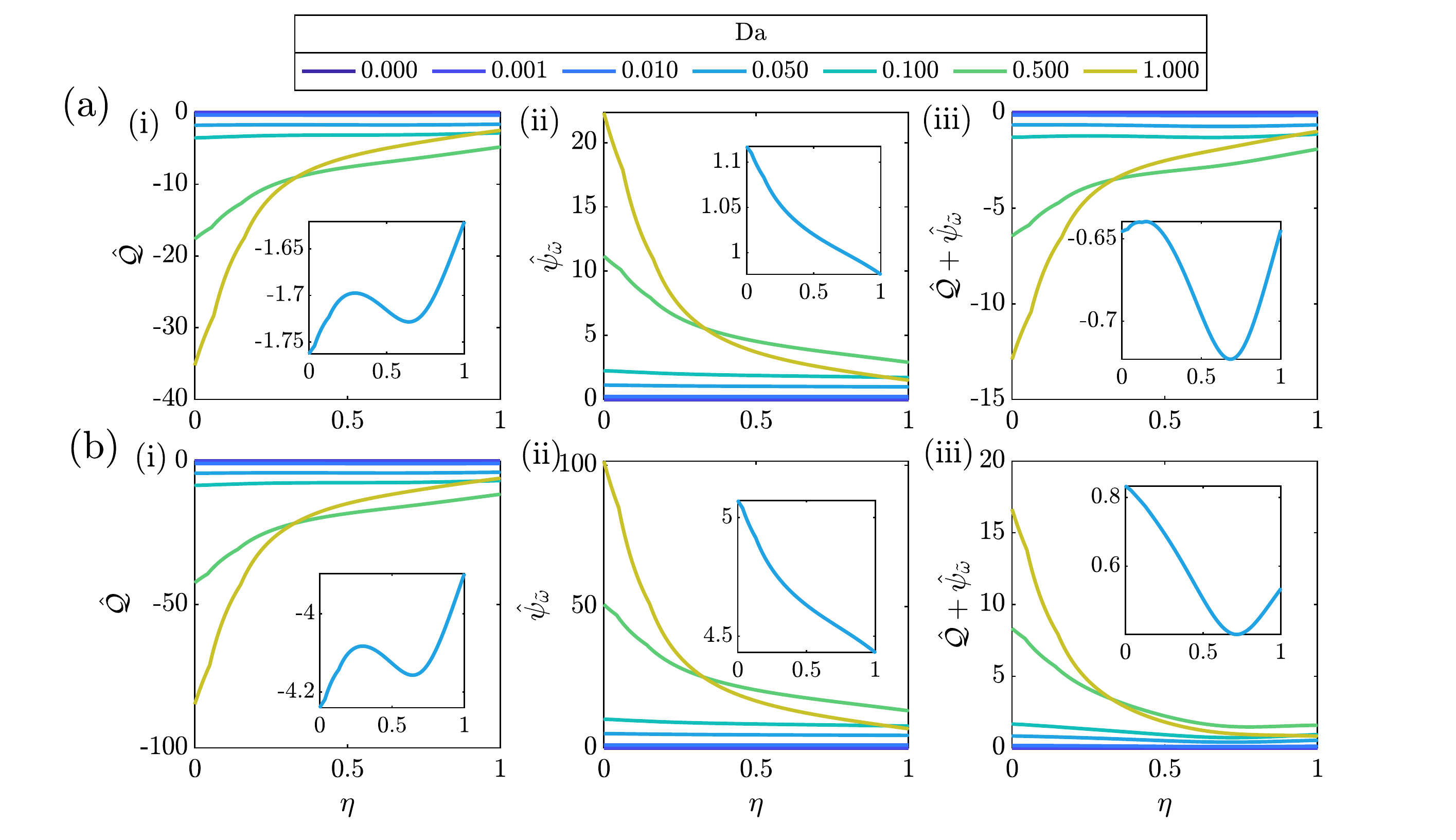}
\caption{
Same quantities as Figure \ref{fig:directnoisefactors_subsonic} for the lin-vel nozzle.
}
\label{fig:directnoisefactors_subsonic_linvelnozz}
\end{figure}
Figures \ref{fig:indirectnoisefactors_subsonic} shows the spatial variations of the different sources of indirect noise ($\S~$\ref{sec:sourcesofnoise}) for He $= 0.5$ \aj{and $\mathrm{Da} = 0 - 10$} {in the CWG nozzle}.
\aj{
The indirect noise factors are constant in the chemically frozen flow (Da $=0$). 
In the case of $\mathrm{Da} \gg 1$, the reaction completes close to the nozzle inlet, which means that the indirect noise factors have a constant magnitude, depending on the compositional properties of the reactants. 
For intermediate values of $\mathrm{Da}$, the indirect noise factors approach the magnitudes for $\mathrm{Da} \gg 1$ downstream in the nozzle, as the reaction approaches completion.
Additionally, the gamma-prime noise factor, $\bar\phi_1$, is a function of the mean flow pressure profile and the heat capacity factor \eqref{eq:phireactass}. Hence, it peaks close to the nozzle throat, and its sign depends on that of $\aleph_1$. The magnitude of $\bar\phi_1$ increases with Da.  
Moreover, the indirect compositional noise factor, $\bar\aleph_1 + \bar\psi_1 + \bar\phi_1$, is a function of compositional noise characteristics of the constituents of reaction depending on their mass fractions (a function of rate of reaction). In hydrogen, we observe that the indirect compositional noise factor decreases with Da.
}
Under the same flow conditions, \aj{for $\mathrm{Da} = 0.05$}, the indirect compositional noise factor, $\bar\aleph_1 + \bar\psi_1 + \bar\phi_1$, is approximately double in the flow with hydrogen inhomogeneities as compared to that with methane (Figure \ref{fig:indirectnoisefactors_subsonic} (a, b)). Likewise, the reacting compositional noise source, $\bar\Theta$ is appoximately ten times larger in hydrogen (Figure \ref{fig:indirectnoisefactors_subsonic} (c, d)). However, it peaks largely around the throat for the {CWG} nozzle geometry, where the flow gradient is maximum.

The direct noise factors of reacting compositional noise are shown in Figure \ref{fig:directnoisefactors_subsonic}. 
\aj{The direct noise factors are zero in a chemically frozen flow ($\mathrm{Da} = 0$), whereas they become constant close to the nozzle inlet for larger Da.
We show the direct noise factors for $\mathrm{Da} = 0.05$ in the insets of Figure \ref{fig:directnoisefactors_subsonic}.
}
On the one hand, the \aj{magnitude of the} heat noise source, $\hat{\mathcal{Q}}$, decreases along the nozzle (Figure \ref{fig:directnoisefactors_subsonic} (a, b) - (i)). The heat noise source, $\hat{\mathcal{Q}}$, remains negative for both cases, which leads to the generation of entropy \eqref{eq:entDirectnoise}. {As explained in $\S~$\ref{sec:sourcesofnoise}, $\hat{\mathcal{Q}}$ is negative because the reaction is exothermic ($\mathcal{Q} > 0$), but the rate of production of fuel  is negative~\eqref{eq:qhat} ($\tilde{\dot{\omega_f^\prime}} < 0$).} 
Because of the large hydrogen chemical energy density, the hydrogen heat noise source is approximately twice as large as that of methane inhomogeneity. 
On the other hand, the reacting chemical potential source, $\hat{\psi}_{\tilde{\omega}}$, remains positive for both cases (Figure \ref{fig:directnoisefactors_subsonic} (a, b) - (ii)), which leads to a decrease in the entropy fluctuations \eqref{eq:entDirectnoise}. The magnitude of $\hat{\psi}_{\tilde{\omega}}$ is approximately five times {larger} for hydrogen as compared to methane.  This is because hydrogen has larger values of the chemical potential noise source, $\bar\psi_1$.
The combined effect of both sources of direct noise is shown in Figure \ref{fig:directnoisefactors_subsonic} (a, b) - (iii). The sum of the direct noise is negative in the case of methane, whereas, it is positive in the case of hydrogen. This physically results in markedly different acoustic behaviour, as shown in the transfer functions ($\S~$ \ref{sec:acousticssubsonic}).  \\ 

The sources of noise in the lin-vel nozzle flow are shown in Figures \ref{fig:indirectnoisefactors_subsonic_linvelnozz} and \ref{fig:directnoisefactors_subsonic_linvelnozz}. 
\aj{We observe largely similar trends with Da as compared to the sources of noise in the CWG nozzle.}
The quantitative and qualitative differences  are caused by different mean-flow properties (Figures \ref{fig:indirectnoisefactors_subsonic},\ref{fig:indirectnoisefactors_subsonic_linvelnozz}, \ref{fig:directnoisefactors_subsonic}, \ref{fig:directnoisefactors_subsonic_linvelnozz}). For example, the heat capacity noise factor, $\bar\aleph_1$, attains a value of $\sim -15$ at the exit in the CWG nozzle (Figure \ref{fig:indirectnoisefactors_subsonic} (a - (i))), whereas it attains $3.5$ in the lin-vel nozzle (Figure \ref{fig:indirectnoisefactors_subsonic_linvelnozz} (a - (i))), in the flow with a methane inhomogeneity \aj{and $\mathrm{Da} = 0.05$}. 
This is because the spatial variation of the fluctuations in the mass fraction of the species is different in the two nozzles for the same Damk\"{o}hler number. 
Owing to the geometry, there is a sharp change in the pressure near the throat in the CWG nozzle, which affects both $\bar\phi_1$ and $\bar\Theta$ (Figures \ref{fig:indirectnoisefactors_subsonic} (c, d),  \ref{fig:indirectnoisefactors_subsonic_linvelnozz} (c, d)). Similar conclusions can be drawn for the direct noise factors (Figures \ref{fig:directnoisefactors_subsonic},\ref{fig:directnoisefactors_subsonic_linvelnozz}). 
 
In conclusion, the sources of noise depend on the reaction chemistry, properties of constituents (reactants and products) of the reaction, their relative amount, mean flow properties. 
The sources of noise affect the acoustic transfer functions, which is discussed in $\S~$ \ref{sec:acousticssubsonic}.

\subsection{Acoustic transfer functions}\label{sec:acousticssubsonic}
Figure \ref{fig:effect_of_fuel_conc_subsonic_ch4h2} shows the acoustic transfer functions {in the CWG nozzle profile}. For small values of Da, the response is close to that of a chemically frozen flow (Da $= 0$).
{The trend in the Damk\"{o}hler number is quantitatively different for methane and hydrogen.}
The magnitudes of transfer functions for a chemically frozen flow are approximately ten times larger in the case of hydrogen, as compared to methane. For instance, $R_{\xi_{{\text{CH}_4}}} = 0.01$ and $R_{\xi_{\text{H}_2}} = 0.1$ for $\mathrm{He} \approx 0.1$. This is because of the large difference in the compositional noise factors. {A larger compositional noise factor results in a larger magnitude of the transfer functions \citep{jain2022compositional}.} 
{In the chemically frozen flow, Da $= 0$, a negligible magnitude of transfer functions is observed for a compact nozzle, He $= 0$. This is because, in a subsonic flow, the divergent section has an adverse pressure gradient. 
In a compact nozzle, the acoustic waves generated in the convergent section are cancelled out by those generated in the divergent section \citep{duran2013solution}. However, in a non-compact nozzle, the increase in Helmholtz number generates a phase difference between these waves, which manifests itself as larger acoustic waves (Figure \ref{fig:effect_of_fuel_conc_subsonic_ch4h2} (a, b, e, f)). For the reacting flow, we obtain non-zero values of the transfer functions in a compact nozzle. This is because the chemical reactions introduce an additional phase shift. 
} \\ 

For methane, the magnitude of the reflection coefficient increases with the value of Da $> 0.5$. However, a non-monotonic behavior is observed for the smaller Damk\"{o}hler numbers as shown in Figure \ref{fig:effect_of_fuel_conc_subsonic_ch4h2} (a). The transmission coefficient increases with Da, and starts decreasing for He $\gtrapprox 0.1$ in the analysed flow (Figure \ref{fig:effect_of_fuel_conc_subsonic_ch4h2} (b)). It decreases to values lower than that in a chemically frozen case.
On the other hand, for hydrogen, the magnitude increases with Da (Figure\ref{fig:effect_of_fuel_conc_subsonic_ch4h2} (e)), whereas the magnitude of the transmission coefficient largely increases with Da (Fig \ref{fig:effect_of_fuel_conc_subsonic_ch4h2} (f)). However, the effect of Da on the magnitude of the transmission coefficient becomes smaller with an increase in the Helmholtz number.  
Figure \ref{fig:effect_of_fuel_conc_subsonic_ch4h2} (c,d, g, h) shows how the phase of the transmitted and reflected waves changes with the Damk\"{o}hler number, Da, and Helmholtz number, He.
The acoustic transfer functions of the two fuels are different because of two reasons. 
First, the {nature of the }chemical reaction {and the properties and mass fractions of the products are different}.
For instance, the heat released from  hydrogen is approximately twice as large as that of methane. This means that the same amount of hydrogen produces a  larger amount of $\text{H}_2\text{O}$ as compared to methane, which results in a large difference in values of sources of compositional noise (Table \ref{tab:sources}) as shown in Figures \ref{fig:indirectnoisefactors_subsonic} and \ref{fig:directnoisefactors_subsonic}.
Second, different Damk\"{o}hler numbers and Helmholtz numbers affect the phase of the acoustic waves. Physically, some Helmholtz number and Damk\"{o}hler numbers combinations can result in the net cancellation of the resulting acoustic waves generated in the converging and diverging sections. \\ 

Figure \ref{fig:effect_of_fuel_conc_subsonic_ch4h2_linvelnozz} shows the variation of the transfer functions {in the lin-vel nozzle}. The magnitudes of the transfer functions are approximately ten times larger than those observed in the {CWG nozzle}.
For methane, we observe a similar trend {in the lin-vel nozzle} for the reflection coefficient {as compared to CWG nozzle}.  On the other hand, the transmission coefficient for the {lin-vel nozzle} increases with Da.
For hydrogen, the magnitude of both the reflection and the transmission coefficients increase with Da (Figure \ref{fig:effect_of_fuel_conc_subsonic_ch4h2} (e, f)). The transfer functions of the two nozzles are different because { of the different mean flow properties and sources of noise, as explained in $\S~$\ref{sec:son_subsonic}.}

It can be concluded that the response of the nozzle depends on the combined effect of mean flow, hence the nozzle geometry, and the Damk\"{o}hler number, Da, and the Helmholtz number, He. This combined interaction affects the phase and magnitudes of the reflected waves in a subsonic flow, which are the key quantities for noise emission and stability.

\begin{figure} 
\centering    
\includegraphics[width=1\textwidth]{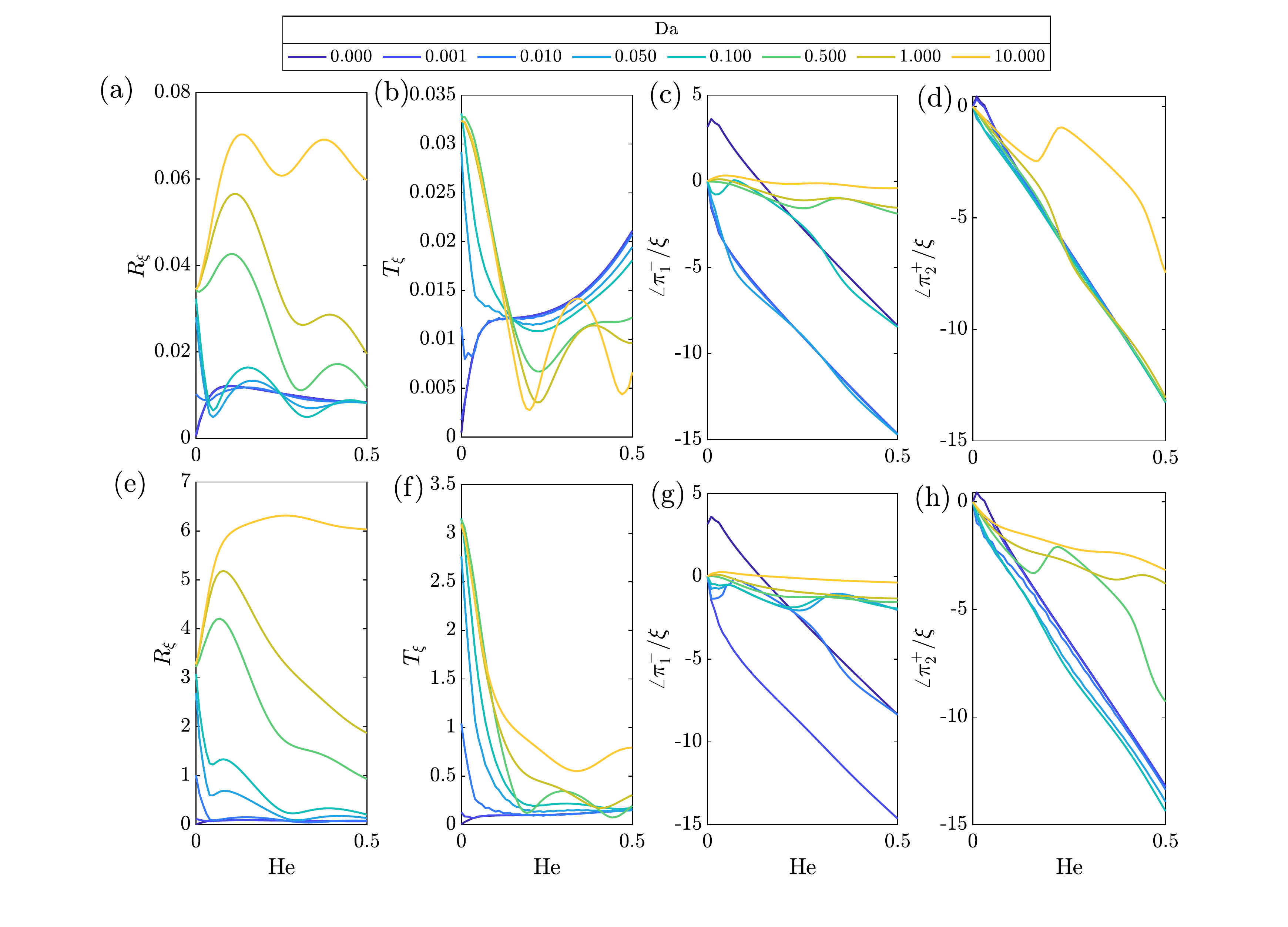}
\caption{Compositional-acoustic (a, e) reflection coefficient, (b, f) transmission coefficient, (c, g) phase of the reflected acoustic wave, (d, h) phase of the transmitted acoustic wave for a reacting mixture of air and (a - d) methane, (e - h) hydrogen in a subsonic nozzle flow ({CWG nozzle}) with throat Mach number $M_t = 0.6$.}
\label{fig:effect_of_fuel_conc_subsonic_ch4h2}
\end{figure}
\begin{figure} 
\centering    
\includegraphics[width=1\textwidth]{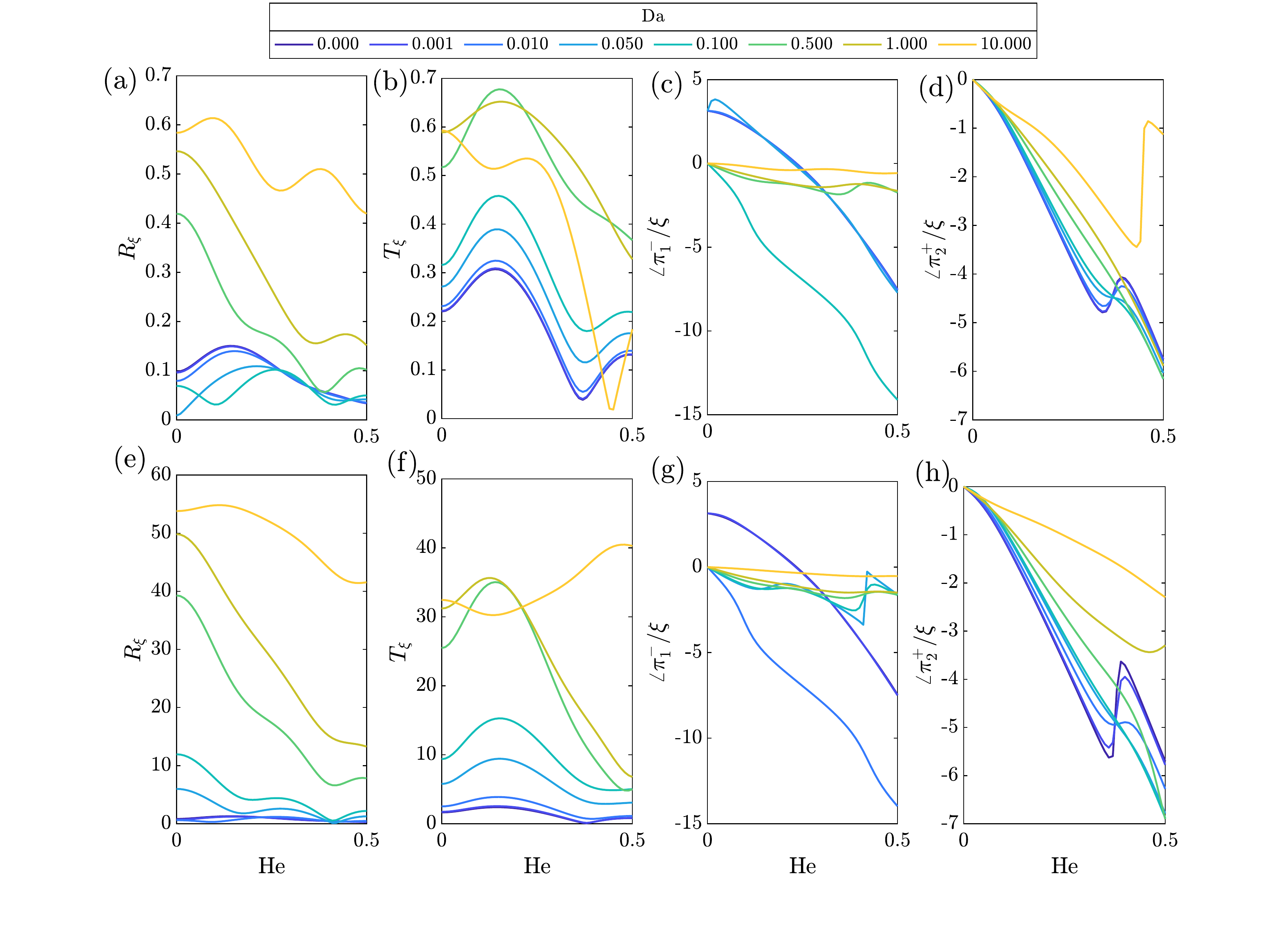}
\caption{Same quantities as Figure \ref{fig:effect_of_fuel_conc_subsonic_ch4h2} for lin-vel nozzle with $M_t = 0.7$.}
\label{fig:effect_of_fuel_conc_subsonic_ch4h2_linvelnozz}
\end{figure}

\FloatBarrier

\section{Acoustic transfer functions in supersonic flows}\label{sec:supersonicreact}
\FloatBarrier
\begin{figure}
\centering    
\includegraphics[width=0.8\textwidth]{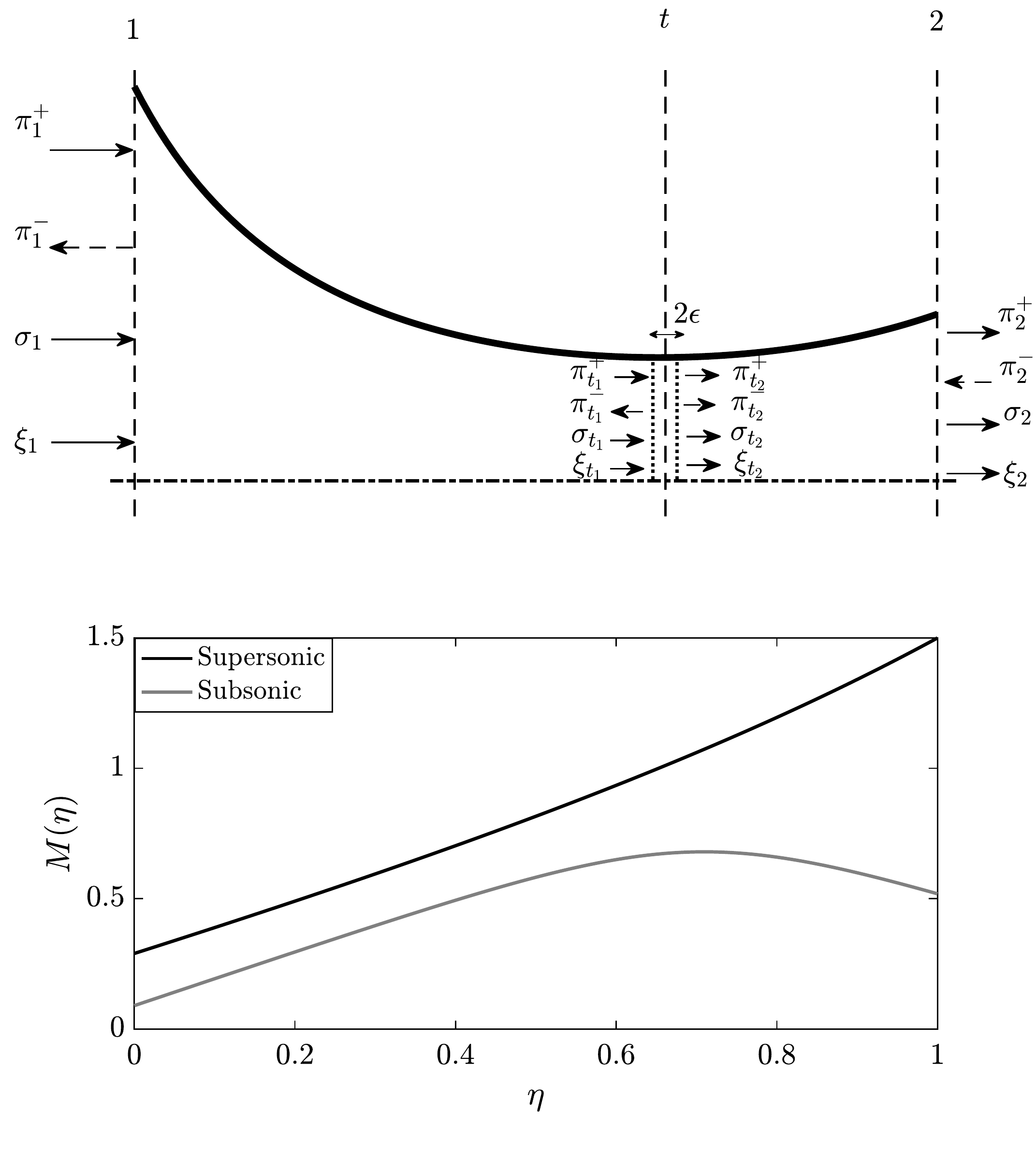}
\caption{(Top) {Lin-vel} nozzle profile. (Bottom) Mach number in the nozzle with steady linear velocity profile in a supersonic flow. $(M_{1_\text{sup}} = 0.29, M_{1_\text{sup}} = 1.5)$, and subsonic flow $(M_{1_\text{sub}} = 0.09, M_{t_\text{sub}} = 0.7)$}
\label{fig:linvelnozz_mach}
\end{figure}
\begin{figure}
\centering    
\includegraphics[width=1\textwidth]{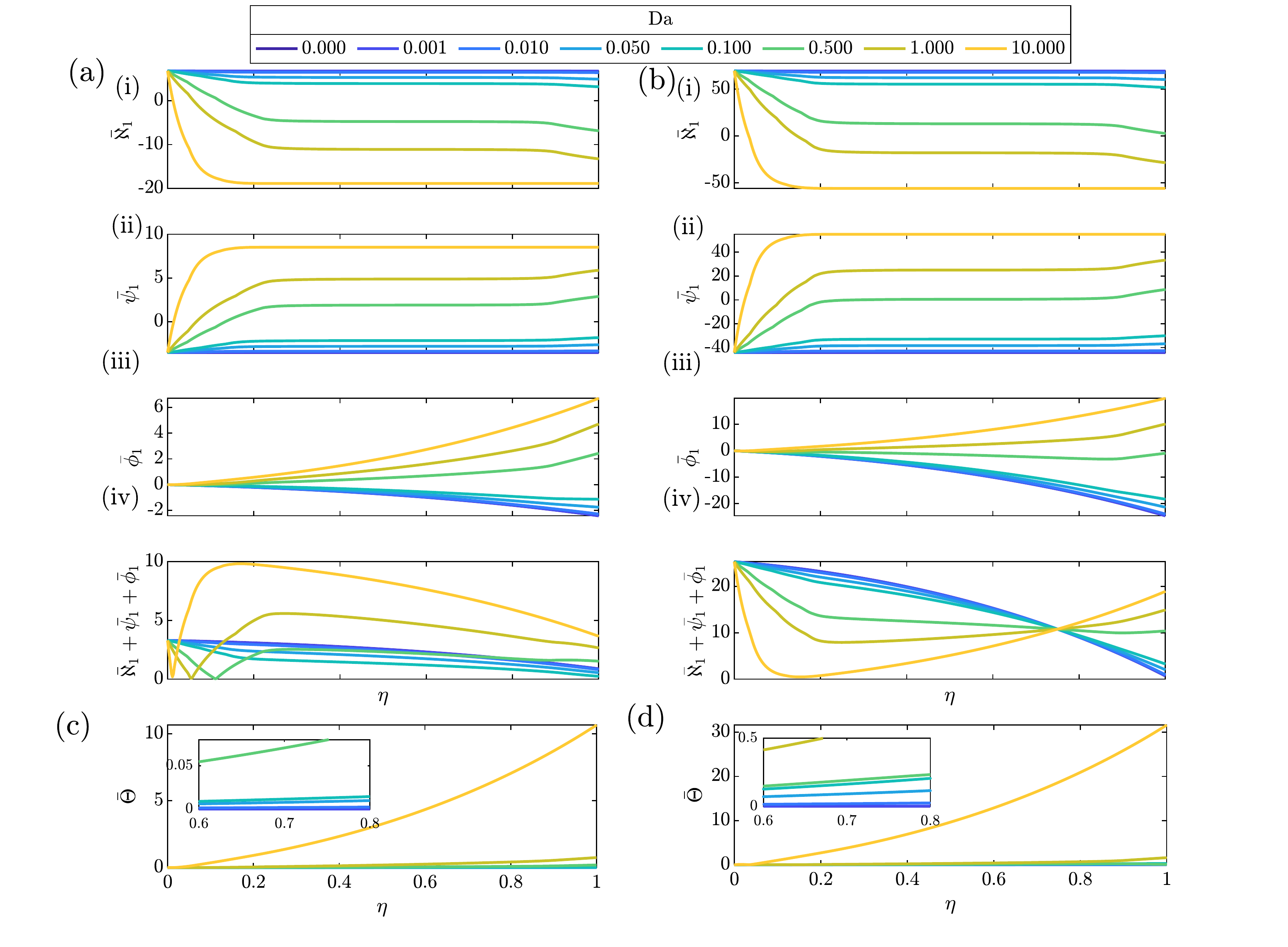}
\caption{
Same quantities as Figure \ref{fig:indirectnoisefactors_subsonic} in a supersonic flow through lin-vel nozzle.
}
\label{fig:psiplusaleph_supersonic}
\end{figure}
\begin{figure}
\centering    
\includegraphics[width=1\textwidth]{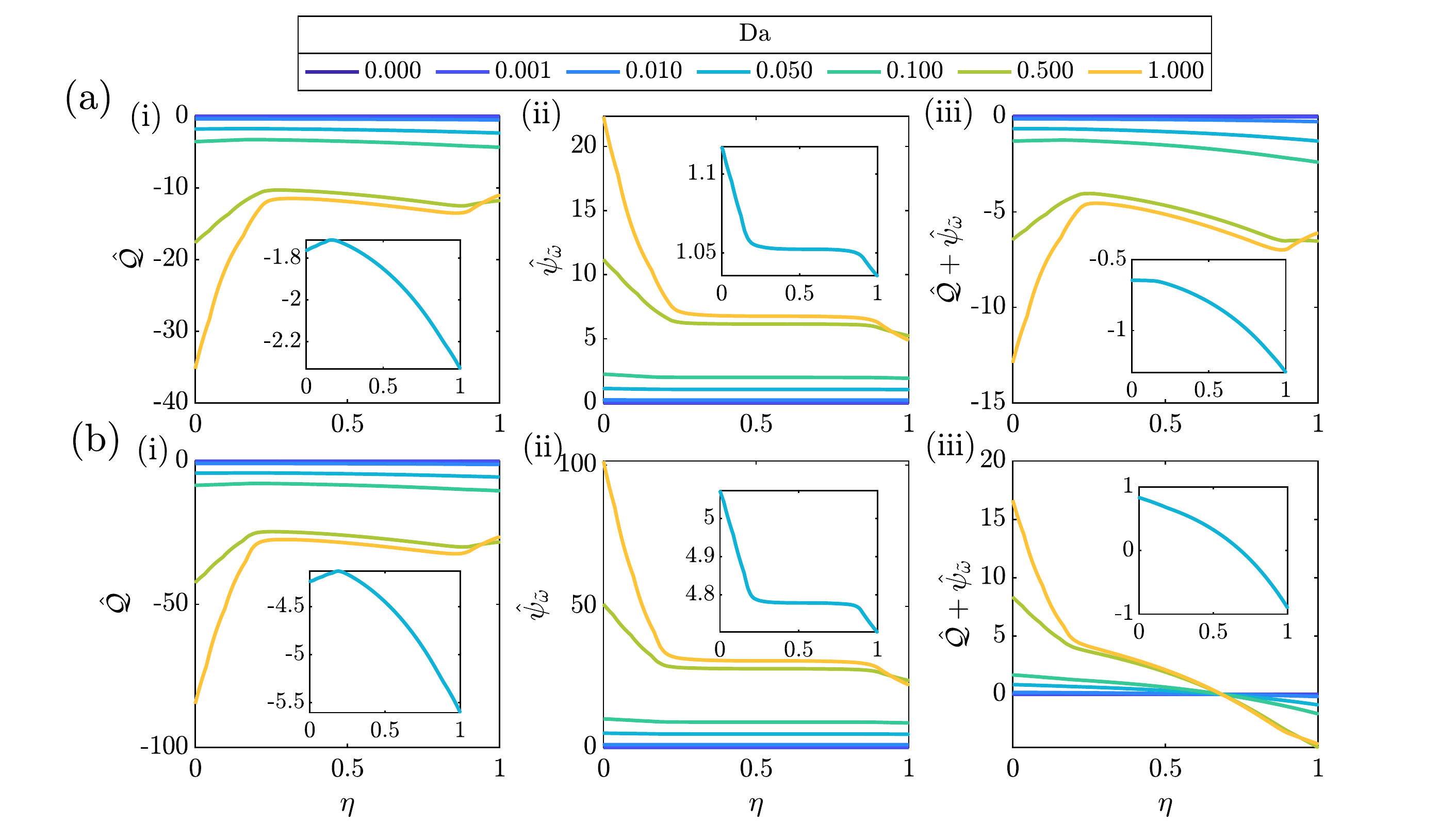}
\caption{
Same quantities as Figure \ref{fig:indirectnoisefactors_subsonic} in a supersonic flow through the lin-vel nozzle.
}
\label{fig:directnoisefactors_supersonic}
\end{figure}
\begin{figure}
\centering    
\includegraphics[width=1\textwidth]{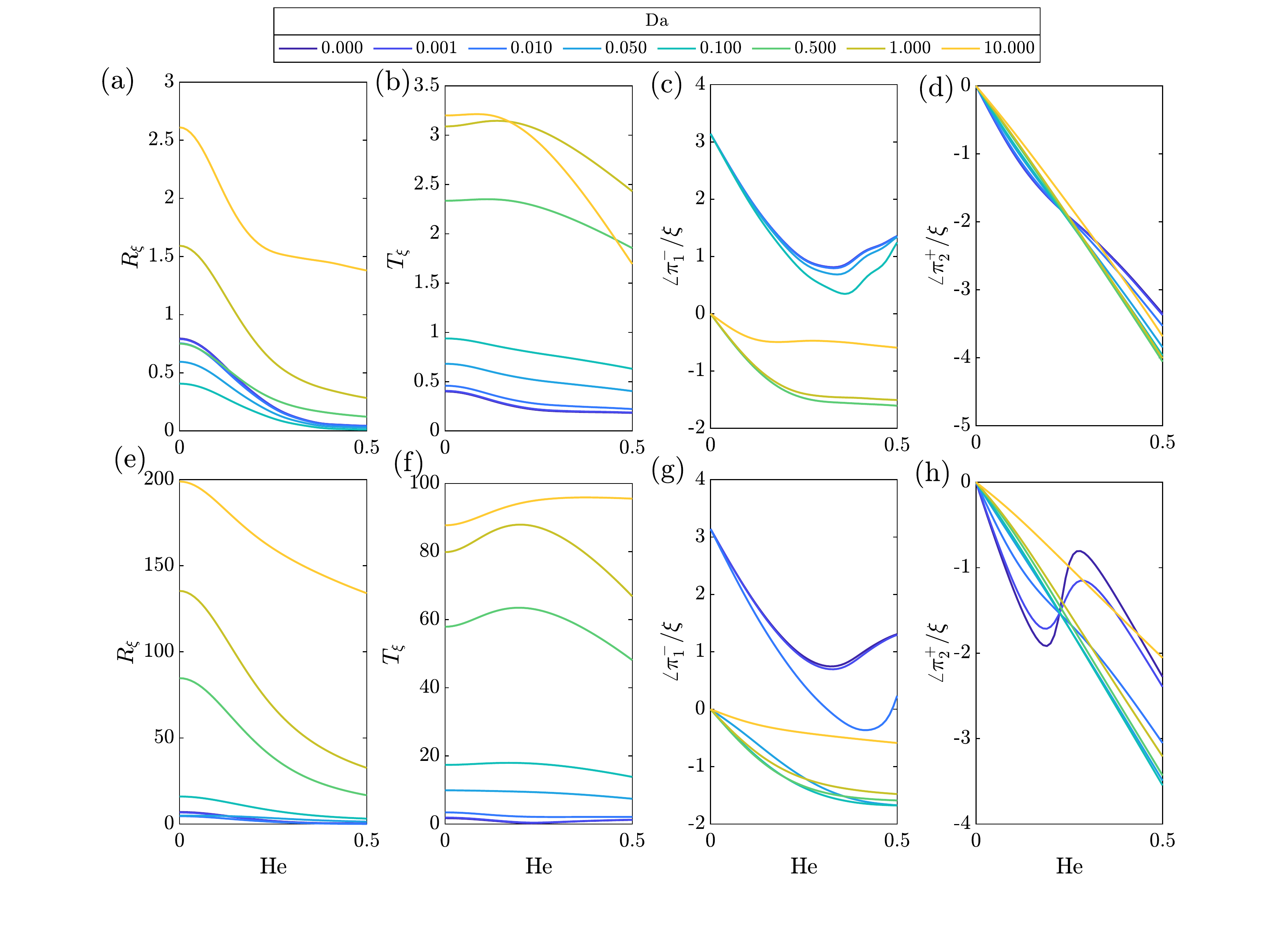}
\caption{Same quantities as Figure \ref{fig:effect_of_fuel_conc_subsonic_ch4h2} for a supersonic flow in lin-vel nozzle.}
\label{fig:effect_of_fuel_conc_supersonic_new}
\end{figure}
In this section, we extend the model to a supersonic flow regime.
The analysis is performed for a linear mean-flow velocity profile ({lin-vel} nozzle profile analysed in $\S~$\ref{sec:subsonicrect}) with the inlet and outlet Mach numbers of $0.29$ and $1.5$, respectively \citep{magri2017indirect}, as shown in Figure \ref{fig:linvelnozz_mach}.
In a supersonic flow, the upstream acoustic wave changes direction at the throat \citep{duran2013solution}. To tackle the singularity, the analysis is performed separately for the convergent and divergent sections with a jump condition at the nozzle throat as shown in Figure \ref{fig:linvelnozz_mach} (top)\begin{align}
    2\frac{u^\prime}{\bar u} + \frac{p^\prime}{\bar\gamma\bar p} (1 - \bar\gamma) - \frac{s^\prime}{\bar c_p}- \sum_{i=1}^N\left(\bar\aleph_{1,i} + \bar\psi_{1,i}\right) Y_i^\prime = 0.
\end{align}
This condition is obtained by imposing $M^\prime/\bar M = 0$ at the nozzle throat \citep{magri2017indirect, jain2022compositional} because there are no fluctuations in the mass fraction and Mach number in a choked flow at the throat.

\subsection{Sources of noise}\label{sec:son_sup}

The sources of indirect noise in a supersonic flow with He $= 0.5$ are shown in Figure \ref{fig:psiplusaleph_supersonic}.
\aj{
The heat capacity factor, $\bar\aleph_1$, and chemical potential function, $\bar\psi_1$ have a similar trend to those in a subsonic regime.
However, unlike the subsonic flow, the flow acceleration increases from the convergent section to the divergent section, resulting in an increase in the pressure throughout. Hence, the magnitude of gamma-prime noise factor, $\bar\phi_1$, monotonically increases throughout the flow and becomes approximately five times larger than that of the subsonic flow. {
Likewise, the reacting source of indirect noise, $\bar\Theta$, increases throughout the nozzle flow (Figure \ref{fig:psiplusaleph_supersonic} (c, d)). The magnitude of $\bar\Theta$ is approximately five times larger than that of the subsonic flow (Figure \ref{fig:indirectnoisefactors_subsonic_linvelnozz} (c, d)).
}
}
As in the subsonic case, the reacting indirect source of compositional noise, $\bar\Theta$ for hydrogen inhomogeneities are larger than those of methane (Figure \ref{fig:psiplusaleph_supersonic} (c, d)).  
 {Because the pressure decreases along the nozzle, in contrast to the subsonic case, the compositional source of indirect noise decreases with Da approximately up to the nozzle throat and increases downstream.}

\aj{The sources of direct noise in a supersonic flow with He $= 0.5$ are shown in Figure \ref{fig:directnoisefactors_supersonic}. The sources for $\mathrm{Da} = 0.05$ are shown in the insets of Figure \ref{fig:directnoisefactors_supersonic}.
}
The heat noise source, $\hat{\mathcal{Q}}$, \aj{is directly proportional to the rate of reaction and inversely proportional to the temperature (Table \ref{tab:sources}). Differently from the subsonic flow, the magnitude of $\hat{\mathcal{Q}}$ increases for smaller Da ($\mathrm{Da} < 0.1$) and decreases for larger Da. This is because, in a supersonic flow temperature monotonically increases, and the rate of reaction decreases along the flow \eqref{eq:omegayf}. With a larger Da, the effect of the  reaction rate dominates, therefore, the magnitude of  $\hat{\mathcal{Q}}$ decreases (see \eqref{eq:Dadef}, \eqref{eq:omegayf} and \eqref{eq:qhat}).
This implies that, in contrast to the subsonic flow, the entropy generation from the heat released in the chemical reaction increases or decreases along the nozzle depending on the Da.
Additionally, in the analysed flow, the magnitudes of both, the heat noise source, $\hat{\mathcal{Q}}$, and the reacting chemical potential source, $\hat{\psi}_{\tilde{\omega}}$ increase with Da.
}
{
Similarly to the subsonic flow, the magnitude of $\hat{\mathcal{Q}}$ for hydrogen is approximately twice as large as that of methane. On the one hand, the heat of reaction generates entropy perturbations for both fuels because $\hat{\mathcal{Q}} < 0$  \eqref{eq:entDirectnoise}.
On the other hand, the reacting chemical potential source, $\hat{\psi}_{\tilde{\omega}}$ dampens the entropy fluctuations ($\hat{\psi}_{\tilde{\omega}} > 0$). 
}
As observed in the subsonic flow ($\S~$\ref{sec:son_subsonic}), there is competition between these two sources of direct noise. 
For methane, the net effect of both direct noise sources is to generate entropy fluctuations ($\hat{\mathcal{Q}} + \hat{\psi}_{\tilde{\omega}} < 0$, Figure \ref{fig:directnoisefactors_supersonic}a,iii), whereas the  direct noise sources in hydrogen reduce the entropy fluctuations \aj{approximately up to the nozzle throat} ($\hat{\mathcal{Q}} + \hat{\psi}_{\tilde{\omega}} > 0$) and enhance them \aj{downstream} ($\hat{\mathcal{Q}} + \hat{\psi}_{\tilde{\omega}} < 0$, Figure \ref{fig:directnoisefactors_supersonic}b,iii).
The effect of sources of noise on the acoustic transfer functions is explained in the next section ($\S~$\ref{sec:acousticssupersonic}.)

{\subsection{Acoustic transfer functions}\label{sec:acousticssupersonic}}
Figure \ref{fig:effect_of_fuel_conc_supersonic_new} shows the effect of chemically reacting compositional inhomogeneities of methane and hydrogen on sound generation. 
{
Because of its large chemical energy density, the magnitudes of the transfer functions of hydrogen are  larger than those of methane inhomogeneities (Figure \ref{fig:effect_of_fuel_conc_supersonic_new} (a, e) and (b, f)). 
For methane (Figure \ref{fig:effect_of_fuel_conc_supersonic_new} (a)), the magnitude of the reflection coefficients decreases up to Da $\approx 0.1$ and increases with Da. The response for Da = $0.5$ is similar to that of a chemically frozen flow.
For the hydrogen inhomogeneity, the reflection coefficient has a magnitude close to that of a chemically frozen flow for Da $= 0.05$. However, the magnitude increases with the Damk\"{o}hler number for Da $> 0.05$ (Figure \ref{fig:effect_of_fuel_conc_supersonic_new} (e)). 
The transmission coefficient increases with an increase in Damk\"{o}hler number, Da (Figure \ref{fig:effect_of_fuel_conc_supersonic_new} (b, f)) up to Da $\approx 1$. The magnitude is close to that of the chemically frozen case for small values of Da. 
The magnitudes of the transfer functions for the two gases are different  because of two reasons. 
First, the reactions generate different amounts of products, which affect the sources of noise and, hence, the transfer functions.
Second, as observed in figure \ref{fig:effect_of_fuel_conc_supersonic_new} (c, d, g, h), the Damk\"{o}hler number, Da, along with the Helmholtz number, He, affect the phases of both reflected and transmitted waves.
}
The transfer functions are a measure of the magnitudes of the acoustic waves when a unit inhomogeneity wave is forced. For hydrogen, the same mass produces a larger heat and reactants as compared to methane. Therefore, we obtain large values of the transfer functions:
{The same mass of hydrogen inhomogeneity results in a large acoustic wave in a supersonic flow for the same Da as compared to that of methane.
}\\
\FloatBarrier

\section{Conclusions}
In this paper, we propose a low-order model of the sound generated by the acceleration of weakly reacting flow inhomogeneities from first principles. 
We physically identify the sources of  direct and indirect noise, which are associated with chemical reactions of multicomponent flows. 
Chemical reactions affect the acoustics through two mechanisms. First, the chemical reaction of the fuel is exothermic, which leads to the generation of entropy fluctuations. 
Second, the reaction changes the composition of the inhomogeneity, which adds to compositional noise. The properties of the products change compositional noise sources.
We apply the model to single-step irreversible reactions of inhomogeneities of 
natural gas (${\text{CH}_4}$) and  hydrogen ($\text{H}_2$), in which the rate of reaction is parameterized by the Damk\"{o}hler number.
Small Damk\"{o}hler numbers correspond to a nearly chemically frozen case, whereas large Damk\"{o}hler numbers correspond to a fast reaction that takes place at the nozzle inlet. In the latter, the flow inside the nozzle can be approximated as a chemically frozen flow of products \aj{of reaction} with air. 
\aj{We compute and analyse the acoustic transfer functions and the sources of noise for two converging-diverging nozzle profiles in a subsonic flow. 
}
First, we {the sources of noise depend on the reaction chemistry,  properties of the reactants, and the products.} 
Second, the magnitude of transfer functions is markedly larger in the case of hydrogen as compared to methane. This means that weakly reacting hydrogen can produce a large amount of indirect noise. 
The response of the nozzle depends on the combined effect of the mean flow (hence, nozzle geometry), reaction chemistry, the Damk\"{o}hler number and the Helmholtz number.
Third, we extend the model and analysis to supersonic flows. 
The magnitudes \aj{of the acoustic transfer functions} are at least twice as large as those of the subsonic flow. 
In the supersonic flow, hydrogen reaction dampens the entropy fluctuations in the convergent part, whilst generating entropy in the divergent section. 
Fourth, the Damk\"{o}hler number affects both the phase and magnitudes of the  transmitted acoustic waves, which produce noise emissions, and the reflected  waves, which can affect thermoacoustic stability. 
This work opens up new possibilities for the accurate modelling of indirect noise and thermoacoustic stability in aeronautical and power-generation nozzles in multi-physics flows.

\subsection*{Acknowledgements}
{A. J. is supported by the University of Cambridge Harding Distinguished Postgraduate Scholars Programme.  L. Magri gratefully acknowledges
financial support from the ERC Starting Grant PhyCo 949388.
}


\end{document}